\documentclass[preprint,3p,12pt,english]{elsarticle}

\usepackage{graphicx}
\usepackage{subfigure}
\usepackage{float}
\usepackage{color}
\usepackage[toc,page]{appendix}
\usepackage{graphicx}
\usepackage{lineno}
\usepackage{booktabs}

\usepackage{todonotes}
\usepackage{multirow}
\usepackage{url}
\usepackage{bm}

\usepackage{hyperref}
\usepackage[useregional]{datetime2}
\usepackage{bm}
\usepackage{newtxmath,newtxtext,amsmath}

\usepackage{helvet}
\def\Put(#1,#2)#3{\leavevmode\makebox(0,0){\put(#1,#2){#3}}}

\let\oldequation\equation
\let\oldendequation\endequation
\renewenvironment{equation}
                 {\linenomathNonumbers\oldequation}
                 {\oldendequation\endlinenomath}
                 
\journal{Nuclear Instruments and Methods A}

\begin{document}

\begin{frontmatter}
\title{Combined Neyman--Pearson Chi-square: An Improved Approximation to the Poisson-likelihood Chi-square}
  
\author{Xiangpan Ji\corref{cor1}}
\author{Wenqiang Gu}
\author{Xin Qian}
\author{Hanyu Wei}
\author{Chao Zhang\corref{cor2}}

\address{Physics Department, Brookhaven National Laboratory, Upton, NY, USA}

\cortext[cor1]{Corresponding author. Email: xji@bnl.gov}
\cortext[cor2]{Corresponding author. Email: czhang@bnl.gov}

\begin{abstract}
  We describe an approximation to the widely-used Poisson-likelihood 
  chi-square using a linear combination of Neyman's and Pearson's 
  chi-squares, namely ``combined Neyman--Pearson chi-square'' 
  ($\chi^2_{\mathrm{CNP}}$). Through analytical derivations and 
  toy model simulations, we show that $\chi^2_\mathrm{CNP}$ leads to
  a significantly smaller bias on the best-fit model parameters
  compared to those using either Neyman's or Pearson's chi-square. When 
  the computational cost of using the Poisson-likelihood chi-square is 
  high, $\chi^2_\mathrm{CNP}$ provides a good alternative given its 
  natural connection to the covariance matrix formalism.
\end{abstract}


\end{frontmatter}



\section{Introduction}\label{sec:intro}
In high-energy physics experiments, it is often convenient to bin the data into a histogram with $n$ bins.
The number of measured events $M_i$ in each bin typically follows a Poisson distribution with the mean
value $\mu_i({\bm\theta})$ predicted by a set of model parameters ${\bm\theta}=(\theta_1,...,\theta_N)$.
The likelihood function of this Poisson histogram can be written as:
\begin{equation}\label{eq:likelihood}
L({\bm\mu}({\bm\theta}); {\bm M}) = \prod_i^n \frac{e^{-\mu_i}\mu_i^{M_i}}{M_i!} \,.
\end{equation}
A maximum-likelihood estimator (MLE) of ${\bm\theta}$ can be constructed by 
maximizing the likelihood ratio~\cite{likelihood-ratio,likelihood-ratio2}
\begin{equation}\label{eq:likelihood-ratio}
\lambda({\bm\theta}) 
= \frac{L({\bm\mu}({\bm\theta}); {\bm M})}{\max L({\bm \mu^{'}}; {\bm M})}
= \frac{L({\bm\mu}({\bm\theta}); {\bm M})}{L({\bm M}; {\bm M})} \,,
\end{equation}
where the denominator is a model-independent constant that maximizes the likelihood of the data without any restriction on the model\footnote{While the estimation of model parameters ${\bm\theta}$ does not depend on the denominator of the likelihood ratio, the chi-square test statistic constructed in this way, such as that in Eq.~\eqref{eq:chi2_Poisson}, can be used to examine the data-model compatibility with a goodness-of-fit test.}. Maximizing this likelihood ratio is equivalent to minimizing the Poisson-likelihood 
chi-square function~\cite{Cash:1979vz,Baker:1983tu}:
\begin{equation}\label{eq:chi2_Poisson}
\chi^2_\mathrm{Poisson}
= -2\ln\lambda({\bm\theta})
=2\sum_{i=1}^n\left(\mu_i({\bf\bm\theta})-M_i+M_i\ln\frac{M_i}{\mu_i({\bf\bm\theta})}\right).  
\end{equation}
The MLE is commonly used in the high-energy physics, as it is generally an 
asymptotically unbiased estimator, and has the advantage of being
consistent and efficient~\cite{pdg}.

At large statistics, the previous Poisson distribution can be approximated 
by a normal (or Gaussian) distribution with mean $\mu_i({\bm\theta})$ and 
variance $\sigma_i^2 = \mu_i({\bm\theta})$. 
The likelihood then becomes:
\begin{equation}\label{eq:Gauss-likelihood}
L_{\textrm{Gauss}}({\bm\mu}({\bm\theta}); {\bm M}) = \prod_i
\frac{1}{\sqrt{2\pi\mu_i({\bm\theta})}}
\exp\left({-\frac{(\mu_i({\bm\theta}) - M_i)^2}{2\mu_i({\bm\theta})}}\right).
\end{equation}
The Gauss-MLE can be similarly constructed through a likelihood ratio:
\begin{equation}\label{eq:Gausslikelihood-ratio}
\lambda_{\textrm{Gauss}}({\bm\theta}) = \frac{L_{\textrm{Gauss}}({\bm\mu}({\bm\theta}); {\bm M})}{\max L_{\textrm{Gauss}}({\bm \mu^{'}}; {\bm M})} \,,
\end{equation}
where the denominator is the maximum of $L_{\textrm{Gauss}}$ without any restriction on the model, and can be derived by calculating $\partial{L_{\textrm{Gauss}}}/\partial{\mu_i^{'}} = 0$.
Maximizing $\lambda_{\textrm{Gauss}}({\bm\theta})$ is equivalent
to minimizing the Gauss-likelihood chi-square function
\begin{equation}\label{eq:chi2_Gauss}
  \begin{aligned}
\chi^2_\mathrm{Gauss} = -2\ln\lambda_{\textrm{Gauss}}({\bm\theta})
&= \sum_{i=1}\left(
\frac{\left(\mu_i({\bm\theta}) - M_i\right)^2}{\mu_i({\bm\theta})}
+ \ln\frac{\mu_i({\bm\theta})}{\mu_i^{'}}
- \frac{(\mu_i^{'}-M_i)^2}{\mu_i^{'}}
\right), \\
\textrm{with}\quad \mu_i^{'} &= \sqrt{1/4 + M_i^2} - 1/2 \,.  
  \end{aligned}
\end{equation} 
While the Gauss-likelihood chi-square is relatively well-known 
(see e.g.~\cite{comp_teststat, Qian:2014nha})~\footnote{We further provide some relevant formulas for the Gauss-likelihood chi-square in~\ref{sec:appendixD}.}, interestingly, it is not widely used in high-energy physics experiments. Instead, a direct chi-square test statistic, namely the Pearson's chi-square, is constructed through:
\begin{equation}
\chi^2_\mathrm{Pearson} = \sum_i \frac{\left(\mu_i({\bm\theta}) - M_i\right)^2}{\mu_i({\bm\theta})} \,.
\end{equation}
Comparing with Eq.~\eqref{eq:chi2_Gauss}, we see $\chi^2_\mathrm{Pearson}$ 
consists of only the first term in $\chi^2_\mathrm{Gauss}$. These two 
chi-squares become asymptotically equivalent when $M_i$ is large.

In practice, the variance $\sigma_i^2$ is often approximated by the measured 
value $M_i$, which is independent of the model parameters. This leads to 
another popular chi-square test statistic in high-energy physics experiments, namely the 
Neyman's chi-square:
\begin{equation}\label{eq:Neyman-chi2}
\chi^2_\mathrm{Neyman} = \sum_i \frac{\left(\mu_i({\bf\bm\theta}) - M_i\right)^2}{M_i} \,.
\end{equation}

Comparing to the MLE from the Poisson-likelihood chi-square, it is known that the 
estimator of model parameters constructed from Pearson's or Neyman's chi-square leads to biases especially when the large-statistics condition is not met~\cite{Baker:1983tu,stat3, Humphrey:2008xz}.
Despite this shortcoming, both $\chi^2_\mathrm{Pearson}$ and 
$\chi^2_\mathrm{Neyman}$ are commonly used in physics data analysis, 
partly because of their close connection to the covariance-matrix formalism:
\begin{equation}\label{eq:chi2_cov}
\chi^2_\mathrm{cov} = \left( {\bm M - \bm{\mu}({\bf\bm\theta})} \right)^T \cdot V^{-1} \cdot \left( {\bm M - \bm{\mu}({\bf\bm\theta})} \right),
\end{equation}
where $V_{ij} = \mathrm{cov}[\mu_i, \mu_j]$ is the covariance matrix of the 
prediction, and can often be calculated through Monte Carlo methods based on the statistical and systematic uncertainties 
of the experiment prior to the minimization of 
$\chi^2_\mathrm{cov}$. In situations where many nuisance parameters~\cite{pdg} 
are required in the likelihood function $L$ as in Eq.~\eqref{eq:likelihood}, 
the covariance matrix format Eq.~\eqref{eq:chi2_cov} has a natural advantage 
of reducing the number of nuisance parameters, thus leads to a faster 
minimization of the $\chi^2$ function.

One method to remove the bias of the estimator from $\chi^2_\mathrm{Pearson}$ 
is through an iteration of the weighted least-squares fit, where the variance 
in one round of $\chi^2_\mathrm{Pearson}$ minimization is replaced by the 
prediction from the best-fit value in the previous round of iteration~\cite{stat1, stat2, Dembinski:2018ihc}. Several modified chi-square 
test statistics have also been proposed in past literatures to mitigate the 
bias issue. For example, $\chi^2_\mathrm{Gauss}$ defined in 
Eq.~\eqref{eq:chi2_Gauss} is a good replacement of $\chi^2_\mathrm{Pearson}$ 
when the number of measurements is large. Similarly, $\chi^2_{\gamma}$ as 
proposed by Mighell~\cite{chi2-gamma} is a good alternative to 
$\chi^2_\mathrm{Neyman}$ when the number of measurements is large. 
Both $\chi^2_{Gauss}$ and $\chi^2_{\gamma}$, however, still lead to biases 
when the number of measurements is small. Redin proposed a solution by 
including a cubic term in $\chi^2_\mathrm{Neyman}$ and 
$\chi^2_\mathrm{Pearson}$~\cite{phystat-2003}, or by reporting a weighted average of fitting results from $\chi^2_\mathrm{Neyman}$ and 
$\chi^2_\mathrm{Pearson}$~\cite{g-2}.

In this paper, we propose a new method through the construction of a chi-square test statistic  ($\chi^2_\mathrm{CNP}$) with a linear combination of Neyman's and Pearson's chi-squares. As an improved approximation to the 
Poisson-likelihood chi-square with respect to either Neyman's or 
Pearson's chi-square, the $\chi^2_\mathrm{CNP}$ significantly reduces the 
bias while keeping the advantage of the covariance matrix formalism. 
This paper is organized as follows.
The construction of $\chi^2_\mathrm{CNP}$ and its covariance matrix format is described in Sec.~\ref{sec:cnp_construction}. 
Three toy examples are presented in Sec.~\ref{sec:toy_model} to illustrate the features and advantages of $\chi^2_\mathrm{CNP}$.
Finally, we summarize the recommended usage in data analysis of counting experiments in Sec.~\ref{sec:discussion}.  


\section{Combined Neyman--Pearson Chi-square ($\chi^2_\mathrm{CNP}$)}\label{sec:cnp_construction}

The bias in the estimator of model parameters ${\bf\bm\theta}$ using Neyman's or Pearson's chi-square can be traced back to the different $\chi^2$ definitions in approximating the Poisson-likelihood chi-square.
To illustrate this, we start with a simple example. A set of $n$ independent counting experiments were performed to measure a common expected value $\mu$. Each experiment measured $M_i$ events.
The three chi-square functions in this case are~\footnote{The treatment for bins where $M_i=0$ is described in \ref{sec:appendixA}.}:
\begin{equation}\label{eq:chi2_Poisson_simple}
  \begin{aligned}
    \chi^2_\mathrm{Poisson} &= 2\sum_{i=1}^n\left(\mu-M_i+M_i\ln\frac{M_i}{\mu}\right),\\
    \chi^2_\mathrm{Neyman} &= \sum_i^n \frac{\left(\mu - M_i\right)^2}{M_i} \,, \\
    \chi^2_\mathrm{Pearson} &= \sum_i^n \frac{\left(\mu - M_i\right)^2}{\mu} \,.
  \end{aligned}
\end{equation}
$\hat{\mu}$ (the estimator of $\mu$) can be calculated through the minimization of Eq.~(\ref{eq:chi2_Poisson_simple}): $\partial \chi^2/\partial \mu = 0$. We obtain:
\begin{equation} \label{eq:estimator_simple}
\hat{\mu}_\mathrm{Poisson} = \frac{\sum_{i=1}^n M_i}{n} \,, \quad
\hat{\mu}_\mathrm{Neyman} = \frac{n}{\sum_{i=1}^n\frac{1}{M_i}} \,, \quad
\hat{\mu}_\mathrm{Pearson} = \sqrt{\frac{\sum_{i=1}^n M_i^2}{n}} \,.
\end{equation}
Given Eq.~\eqref{eq:estimator_simple}, it is straightforward to show that 
$\hat{\mu}_\mathrm{Neyman} \le \hat{\mu}_\mathrm{Poisson} \le \hat{\mu}_\mathrm{Pearson}$,
where the equal sign is only established when all values of $M_i$ are the same.
Since $\hat{\mu}_\mathrm{Poisson}$ is unbiased in this simple example, we see that $\hat{\mu}_\mathrm{Pearson}$ and $\hat{\mu}_\mathrm{Neyman}$ are biased in the opposite directions. 

We further examine the difference in chi-square values. Assuming that $M_i$ and $\mu$ are reasonably large so that $M_i$ is close to $\mu$, a Taylor expansion of $\chi^2_{\mathrm{Poisson}}$ yields:
\begin{equation}\label{eq:Poisson_taylor}
  \begin{aligned}
  \chi^2_\mathrm{Poisson} &= \sum_{i=1}^n 2\left(\mu-M_i-M_i\ln\left( 1 + \frac{\mu-M_i}{M_i} \right)\right) \\
  &\approx \sum_{i=1}^n \left[\frac{(\mu-M_i)^2}{M_i}-\frac{2}{3}\frac{(\mu-M_i)^3}{M_i^2} + O(\frac{(\mu-M_i)^4}{M_i^3})\right].
  \end{aligned}
\end{equation}
From Eq.~\eqref{eq:Poisson_taylor}, it is straightforward to deduce:
\begin{equation}\label{eq:diff_chi2_Poisson_Neyman}
  \begin{aligned}
  \chi^2_\mathrm{Poisson}-\chi^2_\mathrm{Neyman} &\approx -\sum_i^n\frac{2}{3}\frac{(\mu-M_i)^3}{M_i^2} \,, \\
    \chi^2_\mathrm{Poisson}-\chi^2_\mathrm{Pearson}
  &\approx \sum_i^n\frac{1}{3}\frac{(\mu-M_i)^3}{M_i^2} \,.
  \end{aligned}
\end{equation}
Naturally, we can define a new chi-square function as a linear combination of Neyman's and Pearson's chi-squares:
\begin{equation}
\label{eq:chi2_CNP_a}
\chi^2_\mathrm{CNP}  \equiv \frac{1}{3}\left(\chi^2_\mathrm{Neyman}+2\chi^2_\mathrm{Pearson}\right)
= \sum_{i=1}^n \frac{(\mu-M_i)^2}{3/(\frac{1}{M_i}+\frac{2}{\mu})} \,,
\end{equation}
which is approximately equal to $\chi^2_\mathrm{Poisson}$ up to $O(\frac{(\mu-M_i)^4}{M_i^3})$,
better than either $\chi^2_\mathrm{Neyman}$ or $\chi^2_\mathrm{Pearson}$ alone. In this example,
the estimator $\hat{\mu}$ from minimizing $\chi^2_\mathrm{CNP}$ can be derived as:
\begin{equation}
  \hat{\mu}_\mathrm{CNP} = \sqrt[3]{\frac{\sum_{i=1}^n M_i^2}{\sum_{i=1}^n \frac{1}{M_i}}}
  = \sqrt[3]{\hat{\mu}^2_\mathrm{Pearson} \cdot \hat{\mu}_\mathrm{Neyman}} \,,
\end{equation}
which is the geometric mean of two $\hat{\mu}_\mathrm{Pearson}$ and one
$\hat{\mu}_\mathrm{Neyman}$. Since the bias of $\hat{\mu}_\mathrm{Pearson}$ and
$\hat{\mu}_\mathrm{Neyman}$ are in 
the opposite directions, it is easy to see that $\hat{\mu}_\mathrm{CNP}$ has a 
reduced bias.

More generally, when model parameters and systematic uncertainties are included, the $\chi^2_\mathrm{CNP}$ can be written as: 
\begin{equation}
\label{eq:chi2_CNP}
\chi^2_\mathrm{CNP} = \sum_{i=1}^n \frac{(\mu_i({\bf\bm\theta},{\bf\bm\eta})-M_i)^2}{3/(\frac{1}{M_i}+\frac{2}{\mu_i({\bf\bm\theta},{\bf\bm\eta})})} + \sum_{m=1}^{K}\frac{\eta_m^2}{\sigma_{m}^2} \,,
\end{equation}
where ${\bf\bm\theta}=\{\theta_k|k=1,...,N\}$ are model parameters, and 
${\bf\bm\eta}=\{\eta_m|m=1,...,K\}$ are nuisance parameters representing systematic uncertainties constrained with their corresponding standard deviations ($\sigma_m$). 
As an improved approximation to 
$\chi^2_\mathrm{Poisson}$, $\chi^2_\mathrm{CNP}$ in Eq.~\eqref{eq:chi2_CNP} will naturally lead to a reduced bias in estimating model parameters ${\bf\bm\theta}$,  such as the normalization or the shape of the histograms, than using $\chi^2_\mathrm{Neyman}$ or $\chi^2_\mathrm{Pearson}$.

It is worth noting that in $\chi^2_{\mathrm{CNP}}$, the variance of the Gaussian distribution for the $i$th bin
is approximated as $3/(\frac{1}{M_i}+\frac{2}{\mu_i})$, while for $\chi^2_{\mathrm{Neyman}}$
and $\chi^2_{\mathrm{Pearson}}$ they are $M_i$ and $\mu_i$, respectively. From this we can further deduce
the covariance matrix format of the $\chi^2_{\mathrm{CNP}}$. Following Ref.~\cite{Demortier:CDF},
when $\mu_i$ can be approximated as being linearly dependent on nuisance
parameters: $\mu_i = \mu_i^0+\sum_m^K \eta_m s_{mi}$, the chi-square format with pull terms (e.g.~Eq.~\ref{eq:chi2_CNP}) is
equivalent to the chi-square in the covariance matrix format (Eq.~\ref{eq:chi2_cov}).
In this case, the covariance matrix $V$ can be written as
\begin{equation}
V_{ij} = V_{ij}^\mathrm{stat}+V_{ij}^\mathrm{syst}
\,, \quad V_{ij}^\mathrm{syst} = \sum_m^K \sigma_m^2 s_{mi} s_{mj} \,. 
\end{equation}
Therefore, the covariance matrix format of $\chi^2_{\mathrm{CNP}}$ becomes:
\begin{equation}
\label{eq:cov_CNP}
(\chi^2_{\mathrm{CNP}})_\mathrm{cov} 
= \left( {\bm M - \bm{\mu}}({\bf\bm\theta}) \right)^T \cdot (V^{\mathrm{stat}}_{\mathrm{CNP}}({\bf\bm\theta})+V^{\mathrm{syst}})^{-1} \cdot \left( {\bm M - \bm{\mu}}({\bf\bm\theta}) \right),
\end{equation}  
where
\begin{equation}
\label{eq:Eq_StatError}
V_{\mathrm{CNP}}^{\mathrm{stat}}({\bf\bm\theta})_{ij} \equiv 3/(\frac{1}{M_i}+\frac{2}{\mu_i({\bf\bm\theta})}) \delta_{ij}.
\end{equation}
Note that in Eq.~\eqref{eq:Eq_StatError} we have approximated $\mu_i(\bm\theta, \bm\eta) \approx \mu_i(\bm\theta)$ by
fixing the nuisance parameters at their externally constrained (i.e.~nominal) values. This is necessary because the above derivation requires
that uncertainties must be independent of the nuisance parameters $\bm \eta$~\cite{Demortier:CDF}.

While the biases of Neyman's and Pearson's chi-squares are well-known~\cite{Baker:1983tu,stat3, Humphrey:2008xz}, the construction of $\chi^2_\mathrm{CNP}$ is, interestingly, new. This could be partially caused by the fact that in low-statistics experiments where the use of $\chi^2_{\mathrm{Neyman}}$ or  $\chi^2_{\mathrm{Pearson}}$
leads to a high bias, the Poisson-likelihood chi-square is generally used instead. $\chi^2_{\mathrm{CNP}}$,
however, provides certain advantages in situations where either the number of nuisance parameters is too high,
or the likelihood function is analytically difficult to write. In the next section, we demonstrate the features
and advantages of $\chi^2_\mathrm{CNP}$ with three toy examples of increasing complexity. 
Before that, below we briefly discuss the expected performance of $\chi^2_\mathrm{CNP}$ regarding two other common properties of a test statistic: the \emph{goodness of fit} and the \emph{interval estimation}.

\subsection{Goodness of fit}
\label{sec:goodness-of-fit}
In a goodness-of-fit test, the test statistic (e.g. $\chi^2_{\mathrm{Poisson}}$) 
is evaluated at the estimator $\hat{\mu}$ (i.e.~the best-fit value of $\mu$). Assuming its distribution 
following a chi-square distribution with the corresponding number of 
degrees of freedom, a p-value can be calculated to evaluate the 
compatibility between the data and the model. Although $\chi^2_\mathrm{CNP}$ 
can be used to perform such a test, it does not hold a particular advantage 
over the preferred choice of $\chi^2_{\mathrm{Pearson}}$~\cite{comp_teststat}. As shown in Fig.~\ref{Fig::goodness-of-fit} in Sec.~\ref{sec:example_1}, the 
distributions of $\chi^2_{\mathrm{Poisson}}$, $\chi^2_{\mathrm{Gauss}}$, 
$\chi^2_{\mathrm{Neyman}}$, $\chi^2_{\mathrm{Pearson}}$, and 
$\chi^2_{\mathrm{CNP}}$ all deviate from the ideal chi-square distribution 
at low values of $\mu_\mathrm{true}$, while $\chi^2_{\mathrm{Pearson}}$ deviates the 
least. In addition, the mean of the $\chi^2_{\mathrm{Pearson}}$ distribution
equals to the number of degrees of freedom at all $\mu_\mathrm{true}$'s.
Therefore, following Ref.~\cite{comp_teststat}, we recommend to use 
$\chi^2_{\mathrm{Pearson}}$ together with the least-biased estimator $\hat\mu$ 
(from e.g.~$\chi^2_{\mathrm{Poisson}}$ or $\chi^2_{\mathrm{CNP}}$) to perform 
the goodness-of-fit test.

\subsection{Interval estimation}
\label{sec:CI}
It is well known that the construction of confidence
intervals in the frequentist approach not only depends on the choice of test statistics $T$, but also on its actual procedure. Within the high-energy 
physics community, there are two popular procedures in setting the confidence 
intervals, which we describe below.

The first procedure is based on the Wilks' theorem~\cite{Wilks}. The confidence interval is set by placing a certain threshold $c$ on the distribution of $\Delta T\left(\mu\right) = T\left(\mu\right) - T_{min}$,  where $\mu$, $T\left(\mu\right)$, and $T_{min}$ are the parameter of interest, the test 
statistic evaluated at $\mu$, and the global minimum of the $T\left(\mu\right)$ for all model parameters, respectively. 
Under the conditions that i) the two hypotheses are nested, ii) the parameters of the larger hypothesis (e.g.~$T\left(\mu\right)$) are all uniquely defined in the smaller hypothesis (e.g.~$T_{min}$), and not on the limits of the allowed region, and iii) data are asymptotic, Wilks proves that the negative-two-log-likelihood-ratio test statistic $\Delta T$ follows a chi-square distribution and the estimator $\hat{\mu}$ follows a normal distribution centered around the true value $\mu_\textrm{true}$. 
Consequently, the threshold $c$ can be conveniently calculated. For instance, the threshold $c$ for the 68\%, 95\%, and 99.7\% confidence intervals are 1, 4, and 9, respectively, assuming $\Delta T$ follows a chi-square distribution with one degree of freedom. 
With this procedure, the 
correctness of the confidence interval coverage depends on 
the validity of the Wilks' theorem. 
As demonstrated in 
Eq.~\eqref{eq:diff_chi2_Poisson_Neyman} and Eq.~\eqref{eq:chi2_CNP_a}, 
$\chi^2_{\mathrm{CNP}}$ is an improved approximation to the negative-two-log-likelihood-ratio
of the Poisson distribution (i.e.~$\chi^2_{\mathrm{Poisson}}$), and it leads to a reduced bias in the estimator $\hat\mu$ compared to those from $\chi^2_{\mathrm{Neyman}}$ and 
$\chi^2_{\mathrm{Pearson}}$. Therefore, the conditions of the Wilks' theorem 
are better met with $\chi^2_{\mathrm{CNP}}$, which means the the chi-square distribution is a better approximation to the $\Delta T$ distribution from $\chi^2_{\mathrm{CNP}}$. Fig.~\ref{Fig::delta-chi2} in Sec.~\ref{sec:example_1} shows one such example. Consequently, we expect a more proper coverage of the confidence interval using $\chi^2_{\mathrm{CNP}}$ when compared to those using $\chi^2_{\mathrm{Neyman}}$ or
$\chi^2_{\mathrm{Pearson}}$ under this procedure. 

The second procedure is commonly referred to as the Feldman-Cousins 
approach~\cite{Feldman:1997qc} in the high-energy physics community. 
In this procedure, the construction of the
confidence interval strictly follows a frequentist definition (Neyman construction) with an ordering principle based on the value of the likelihood-ratio test statistic (i.e. $\Delta T\left(\mu\right) = T\left(\mu\right) - T_{min}$ with $T = \chi^2_{\mathrm{Poisson}}$ for counting experiments) to ensure a proper frequentist coverage. Sec.~\ref{sec:example_1} shows an example of this procedure with a toy experiment. 
Similarly, the procedure can be defined with an ordering principle based on other $\Delta T$ test statistics (e.g.~$T = \chi^2_{\mathrm{Neyman}}$, 
$\chi^2_{\mathrm{Pearson}}$, or $\chi^2_{\mathrm{CNP}}$), and the constructed confidence intervals would also have proper 
coverages in general. 
In this case, while all of the coverages are proper, a better test statistic is expected to yield a smaller confidence interval in size (or area, volume). As shown in
Table.~\ref{Tab:SimpleToy} of Sec.~\ref{sec:example_1}, the confidence 
interval constructed using $\chi^2_{\mathrm{CNP}}$ is smaller than those 
using $\chi^2_{\mathrm{Neyman}}$ and $\chi^2_{\mathrm{Pearson}}$. This is 
partially caused by the reduced bias in the estimator $\hat{\mu}$ using $\chi^2_{\mathrm{CNP}}$, as will be further discussed in Sec.\ref{sec:example_1}. 

We should note that there are other procedures to set confidence intervals that are less 
affected by certain properties of the test statistics. For example, since the bias ($\delta \mu$) of an estimator 
$\hat{\mu}$ can be evaluated with a Monte Carlo method, one can define 
an alternative test statistic with $\Delta T'\left(\mu\right) = 
T\left(\mu + \delta \mu\right) - T_{min}$. Naturally, the confidence 
interval constructed using $\Delta T'$ with either the thresholding approach based on the Wilks' theorem or the Feldman-Cousins approach  would be less affected by the bias, and performs better than that of $\Delta T$ at the cost of increased computation.


\section{Performance of $\chi^2_{\mathrm{CNP}}$}\label{sec:toy_model}

In this section, we compare the performance of $\chi^2_{\mathrm{Poisson}}$, 
$\chi^2_{\mathrm{Neyman}}$, $\chi^2_{\mathrm{Pearson}}$, and 
$\chi^2_{\mathrm{CNP}}$ with three toy examples. While we focus on the
issue of bias, we also provide comparison results of the goodness-of-fit test
and the interval estimation in the first example to support the discussion in 
Sec.~\ref{sec:goodness-of-fit} and Sec.~\ref{sec:CI}. For completeness of the discussion, 
we add $\chi^2_{\mathrm{Gauss}}$, which has a similar performance to 
$\chi^2_{\mathrm{CNP}}$ in certain scenarios, to the comparison in the 
first example.

\subsection{Example 1: simple counting}\label{sec:example_1}

\begin{figure}[!htbp]
	\begin{center}
		\includegraphics[angle=0, width=6.8cm] {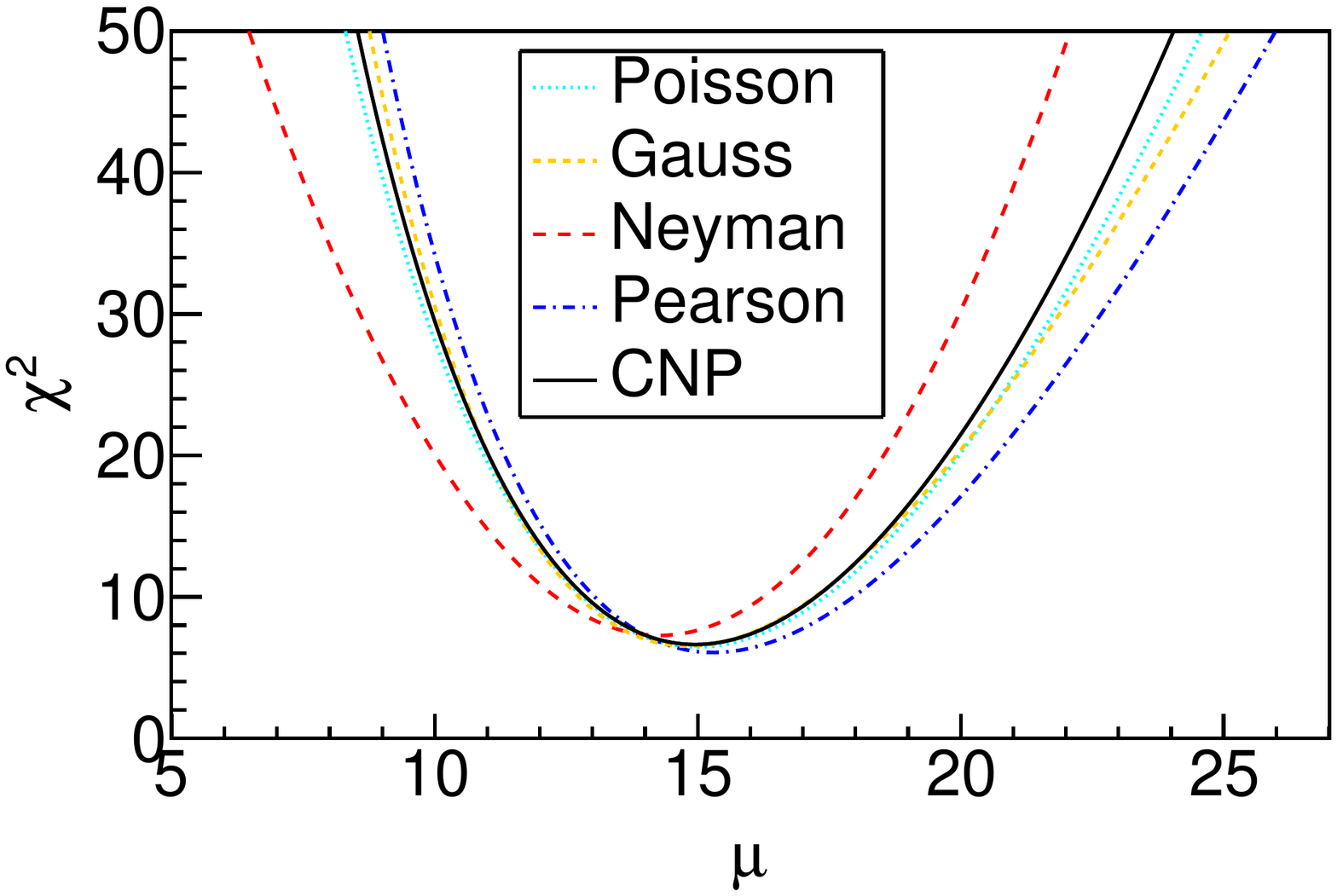}	
		\includegraphics[angle=0, width=6.8cm] {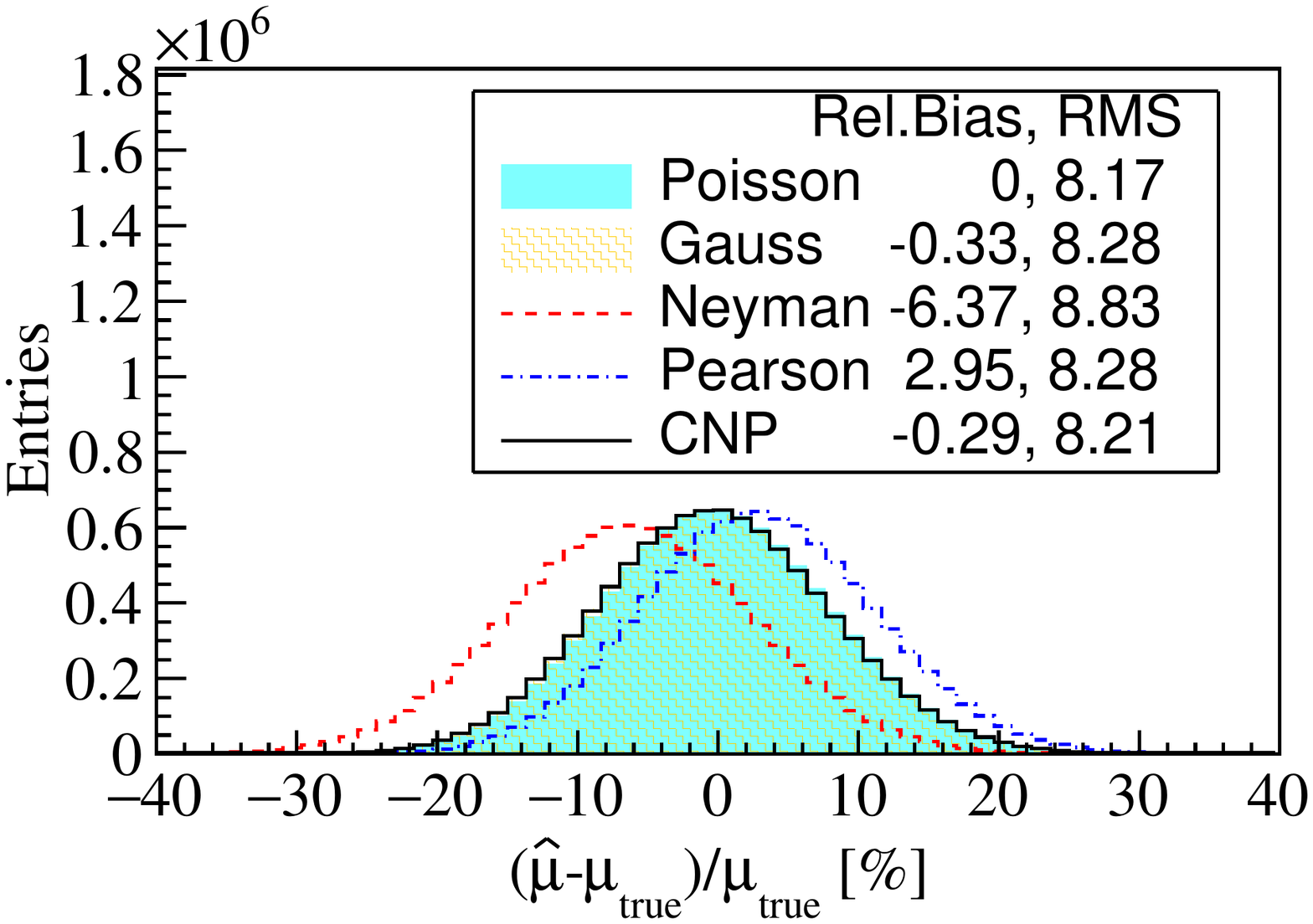}    
		\caption{(Left) The $\chi^2$ curves of the five test statistics:
            $\chi^2_\mathrm{Poisson}$, $\chi^2_\mathrm{Gauss}$, $\chi^2_\mathrm{Neyman}$, 
            $\chi^2_\mathrm{Pearson}$, and $\chi^2_\mathrm{CNP}$
			of one toy experiment with $n=10$ and $\mu_\mathrm{true}=15$. (Right) Distributions of relative difference between $\hat\mu$ and $\mu_\mathrm{true}$ for  
			$\chi^2_\mathrm{Poisson}$, $\chi^2_\mathrm{Gauss}$, $\chi^2_\mathrm{Neyman}$, $\chi^2_\mathrm{Pearson}$ and $\chi^2_\mathrm{CNP}$
			using 10 million toy experiments. The second and third columns in the legend show the relative bias in percentage and the root-mean-square of the relative bias distribution.}
		\label{Fig::SimpleToy}
	\end{center}
\end{figure}

The first example is similar to the one introduced in Sec.~\ref{sec:cnp_construction}. In each toy experiment,
a set of $n$ independent counting measurements were performed to measure a common expected value $\mu$.
The $\chi^2$ curves with $n=10$ and $\mu_\mathrm{true}=15$ of one simulated toy experiment is shown in the left panel of Fig.~\ref{Fig::SimpleToy}.
The minimum location of the $\chi^2$ curve represents the estimator $\hat\mu$. It is clear that $\hat{\mu}_\mathrm{Neyman} < \hat{\mu}_\mathrm{CNP} \approx \hat{\mu}_\mathrm{Poisson} \approx \hat{\mu}_\mathrm{Gauss} < \hat{\mu}_\mathrm{Pearson}$
and the CNP chi-square curve closely resembles the Poisson-likelihood 
chi-square as demonstrated in the previous section.

The relative biases of $\hat\mu$ using $\chi^2_\mathrm{Poisson}$, 
$\chi^2_\mathrm{Neyman}$, $\chi^2_\mathrm{Pearson}$, $\chi^2_\mathrm{Gauss}$
and $\chi^2_\mathrm{CNP}$ are shown in the right panel of Fig.~\ref{Fig::SimpleToy} with 10 million toy experiments. The bias using $\chi^2_\mathrm{Poisson}$ is zero.
The biases using $\chi^2_\mathrm{Neyman}$ and $\chi^2_\mathrm{Pearson}$ have opposite signs. 
The magnitude of mean bias using $\chi^2_\mathrm{Neyman}$ is about twice of that using $\chi^2_\mathrm{Pearson}$.
The bias using $\chi^2_\mathrm{CNP}$ is an order of magnitude smaller than those using $\chi^2_\mathrm{Neyman}$ and $\chi^2_\mathrm{Pearson}$. The bias using $\chi^2_\mathrm{Gauss}$ is similar to $\chi^2_\mathrm{CNP}$.
The variance of $\hat{\mu}_\mathrm{Neyman}$ is notably larger than those of 
the other four test statistics, which are similar. 

\begin{figure}[!htbp]
	\begin{center}
		\includegraphics[angle=0, width=6.8cm] {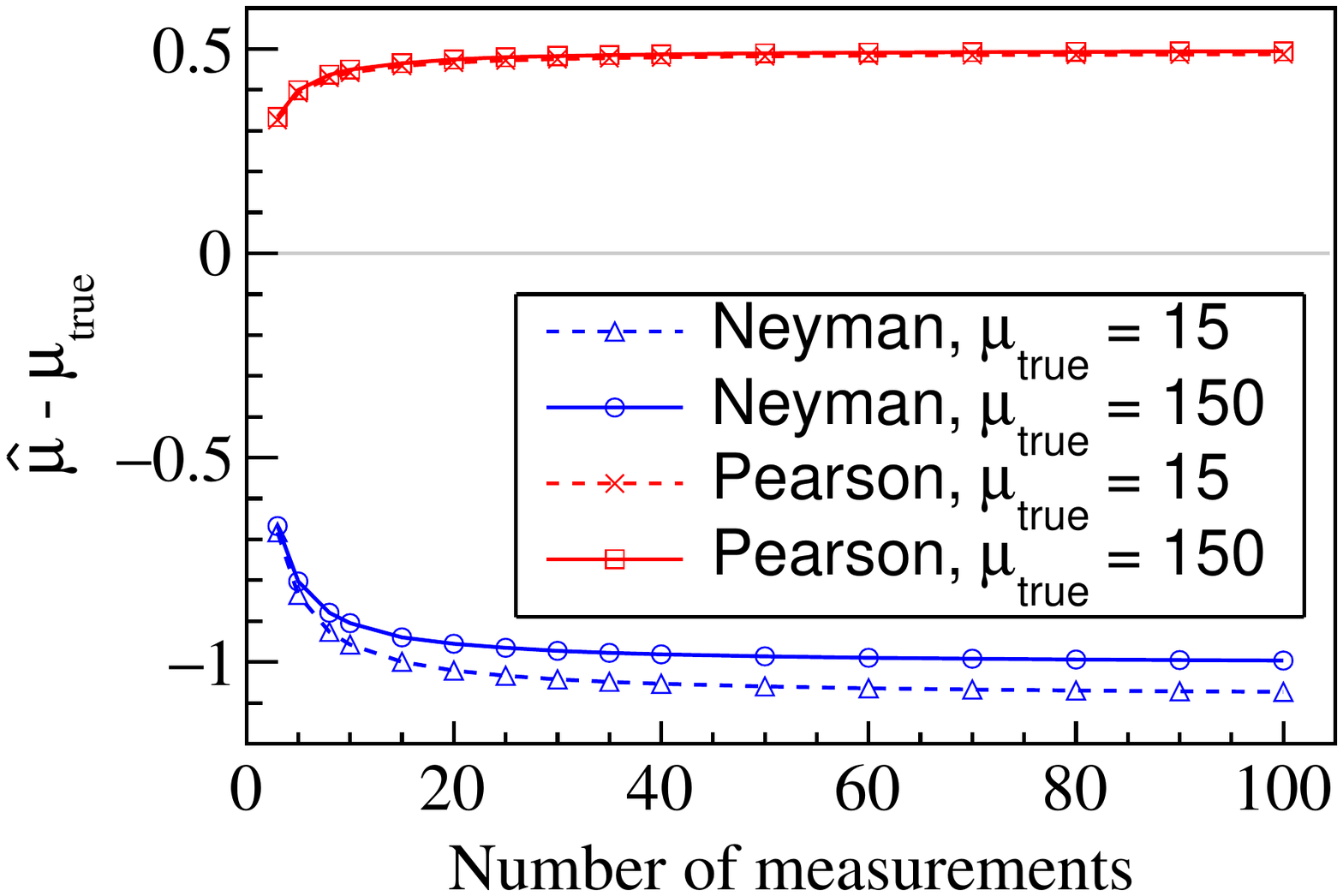}	
		\includegraphics[angle=0, width=6.8cm] {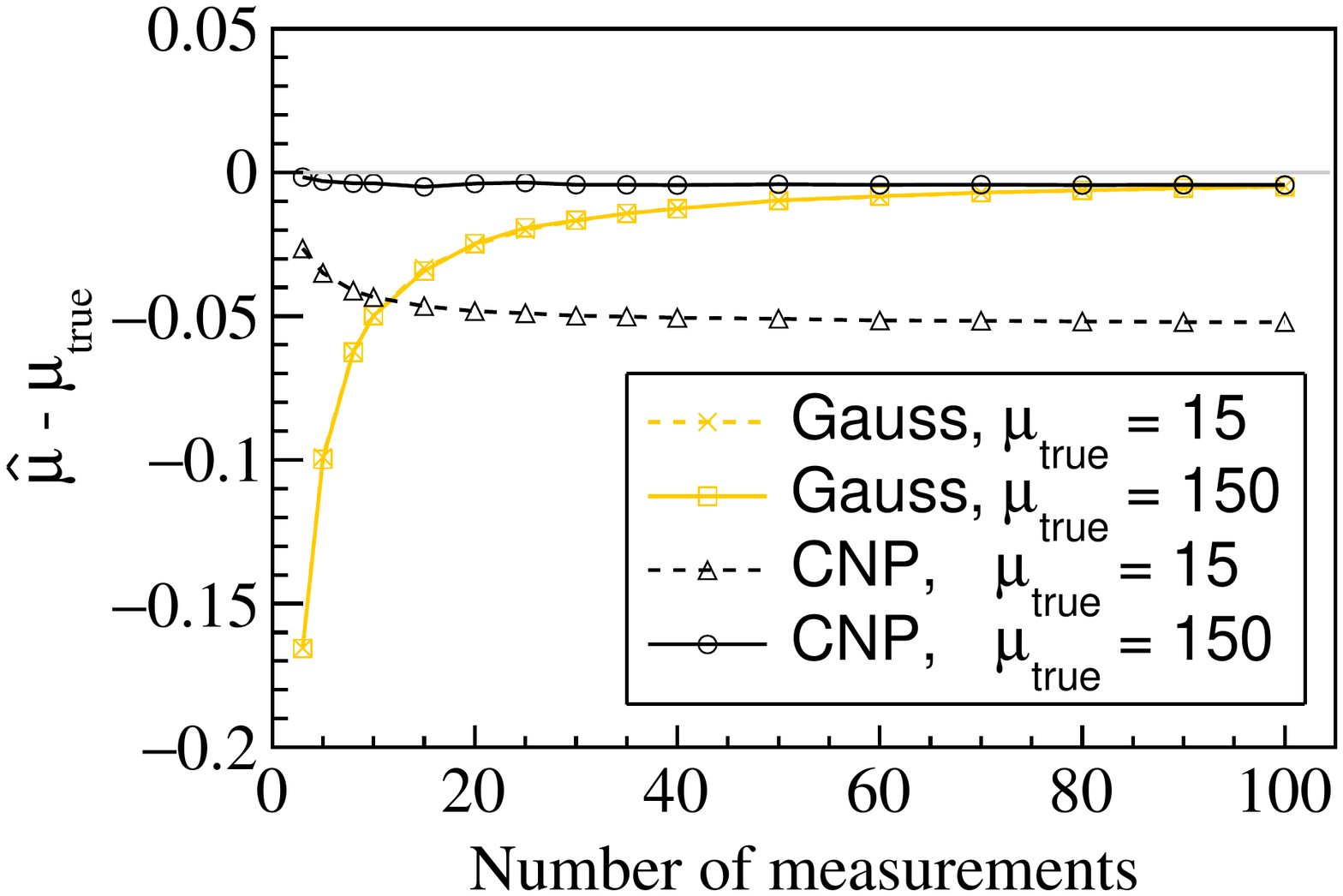}    
		\caption{The absolute biases of $\hat\mu$ as a function of 
the number of measurements $n$ for $\mu_\mathrm{true}=15$ and $\mu_\mathrm{true}=150$. The left panel 
shows the biases using $\chi^2_\mathrm{Pearson}$ and $\chi^2_\mathrm{Neyman}$.
 The right panel shows the biases using $\chi^2_\mathrm{CNP}$ and 
$\chi^2_\mathrm{Gauss}$. Each point is obtained with 10 million toy experiments.}
		\label{Fig::Bias_vs_N}
	\end{center}
\end{figure}

In Fig.~\ref{Fig::Bias_vs_N}, we further study the biases of $\hat\mu$ with different values of $\mu_\mathrm{true}$ and the number of measurements $n$. The biases using 
$\chi^2_{\mathrm{Poisson}}$ are always zero as expected from an unbiased 
estimator in this simple example. The biases using $\chi^2_{\mathrm{Neyman}}$, 
$\chi^2_{\mathrm{Pearson}}$, and $\chi^2_{\mathrm{CNP}}$ become larger as the 
number of measurements $n$ increases. This behavior may not be intuitive, 
but is well known and the proof is provided in \ref{sec:appendixB}. As $\mu$ 
and $n$ increases, the biases of $\hat{\mu}_{\mathrm{Pearson}}$ and 
$\hat{\mu}_{\mathrm{Neyman}}$ approach $1/2$ and $-1$, respectively.     
Beside these observations, the general features of the biases stay the same 
as discussed previously. Most importantly, $\chi^2_{\mathrm{CNP}}$ yields a 
much smaller bias than $\chi^2_\mathrm{Neyman}$ or $\chi^2_\mathrm{Pearson}$ 
in all occasions.

Figure~\ref{Fig::Bias_vs_N} also shows the performance of 
$\chi^2_{\mathrm{Gauss}}$, which is another way to mitigate the bias issue.
Similar to $\chi^2_{\mathrm{CNP}}$, $\chi^2_{\mathrm{Gauss}}$ performs 
much better than $\chi^2_\mathrm{Neyman}$ or $\chi^2_\mathrm{Pearson}$.
We note that the bias of $\hat{\mu}_{\mathrm{Gauss}}$ is less dependent on 
$\mu$, and becomes smaller when $n$ increases. This is expected from the 
central limit theorem, which states that the sum of a large number of 
identically distributed random variables follows a normal distribution. 
Therefore, when the number of measurements is large, $\chi^2_{\mathrm{Gauss}}$
provides a better performance even when $\mu$ is small. On the other hand, 
when number of measurements is not large, $\chi^2_{\mathrm{CNP}}$ shows 
a better performance.

\begin{figure}[!htbp]
	\begin{center}
		\includegraphics[angle=0, width=6.8cm] {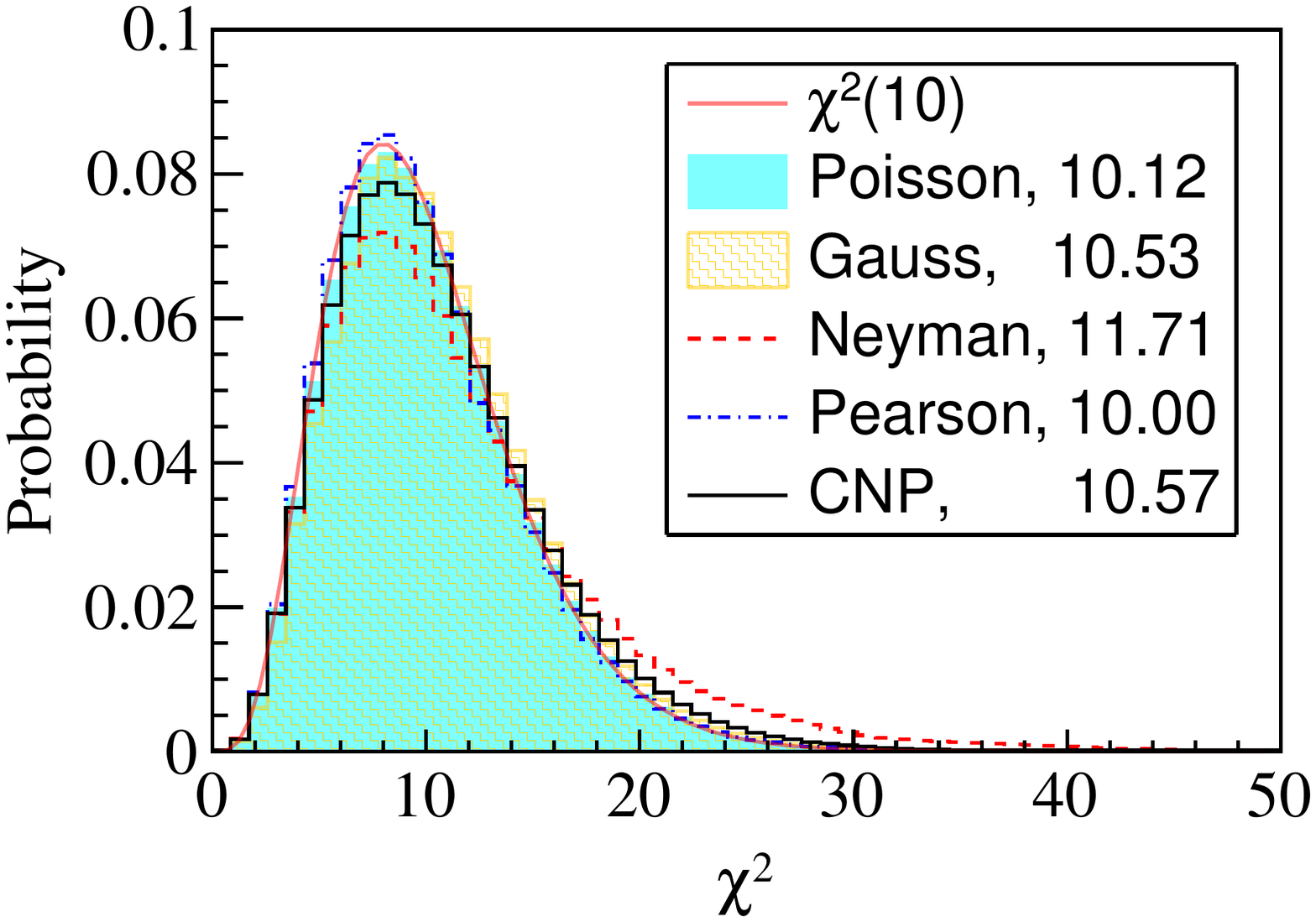}	
		\includegraphics[angle=0, width=6.8cm] {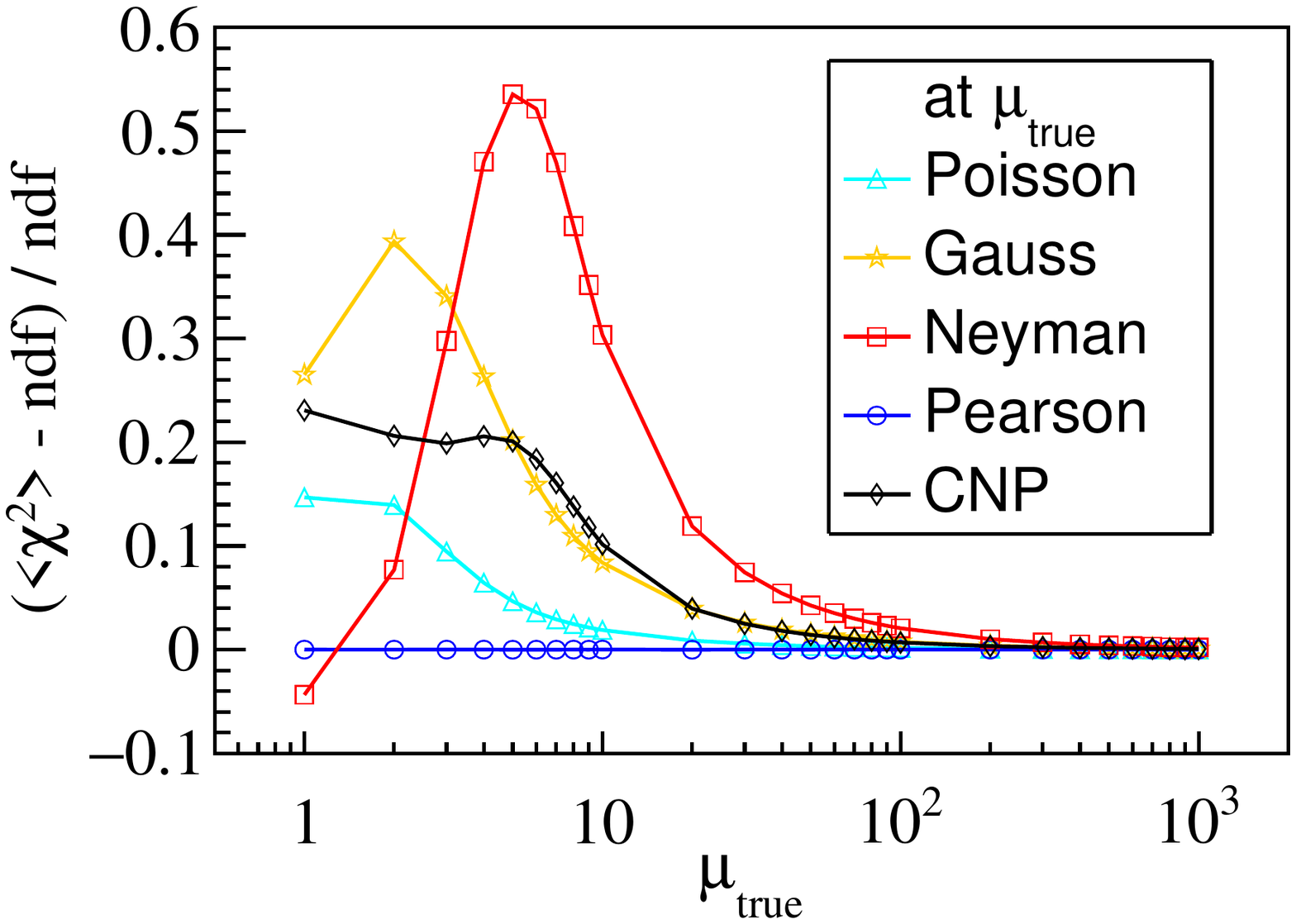}    
		\caption{(Left)  The distribution of the five test statistics:
$\chi^2_\mathrm{Poisson}$, $\chi^2_\mathrm{Gauss}$, $\chi^2_\mathrm{Neyman}$, 
$\chi^2_\mathrm{Pearson}$, and $\chi^2_\mathrm{CNP}$ evaluated at 
$\mu_\mathrm{true}=15$ in the $n=10$ setting using 10 million toy experiments. The ideal chi-square distribution with 10 degrees of freedom, $\chi^2(10)$, is also shown for comparison. The second column in the legend shows the mean of each distribution.
(Right) The relative deviation of the distribution's mean to the number of 
degrees of freedom as a function of 
$\mu_\mathrm{true}$ for these five test statistics. Each point is obtained with 10 million toy experiments in the $n=10$ setting.}
		\label{Fig::goodness-of-fit}
	\end{center}
\end{figure}

Next, we compare the performance on the goodness-of-fit test. The left panel of 
Fig.~\ref{Fig::goodness-of-fit} shows the distribution of the five test 
statistics evaluated at $\mu_\mathrm{true}=15$ in the $n=10$ setting with 10 million toy experiments. The ideal chi-square distribution with 10 degrees of freedom is also shown for 
comparison. All five test statistics deviate from 
the ideal chi-square distribution, with $\chi^2_\mathrm{Pearson}$ being 
the closest and $\chi^2_\mathrm{Neyman}$ deviating the most. 
The mean of $\chi^2_\mathrm{Pearson}$ is exactly 10, and the mean of 
$\chi^2_\mathrm{Neyman}$ is the largest. The right panel of 
Fig.~\ref{Fig::goodness-of-fit} shows the relative deviation of the 
mean to the number of degrees of freedom ($\mathrm{ndf}=10$ in all toy experiments) for the five test 
statistics as a function of $\mu_\mathrm{true}$. It is clear that except for $\chi^2_\mathrm{Pearson}$, the other four test statistics are not ideal in this metric when $\mu_\mathrm{true}$ is less than a few tens, 
with $\chi^2_\mathrm{Neyman}$ being the worst. Ref.~\cite{comp_teststat} provides a good discussion on this behavior. 

\begin{figure}[!htbp]
    \begin{center}
        \includegraphics[angle=0, width=6.8cm] {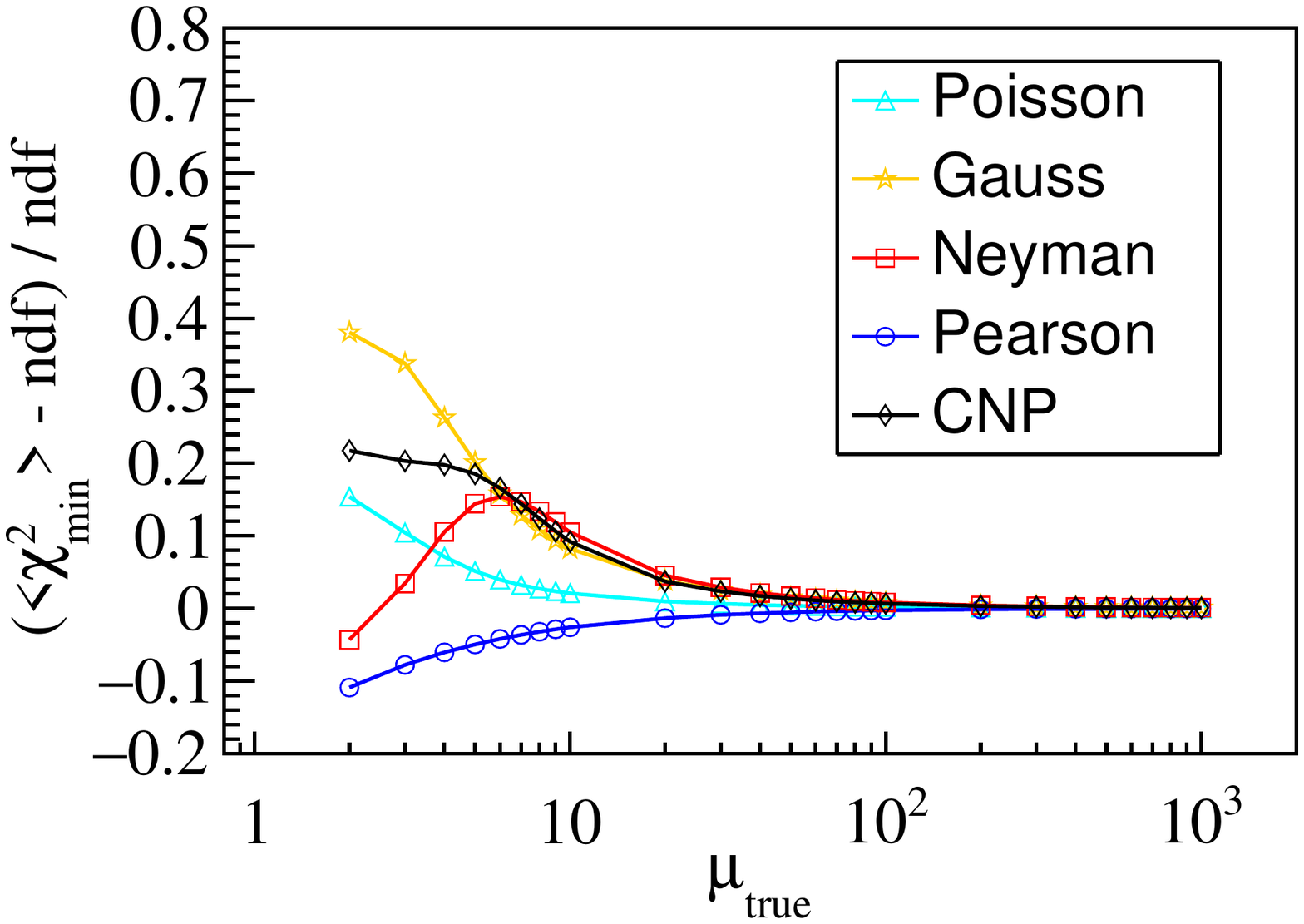}	
        \includegraphics[angle=0, width=6.8cm] {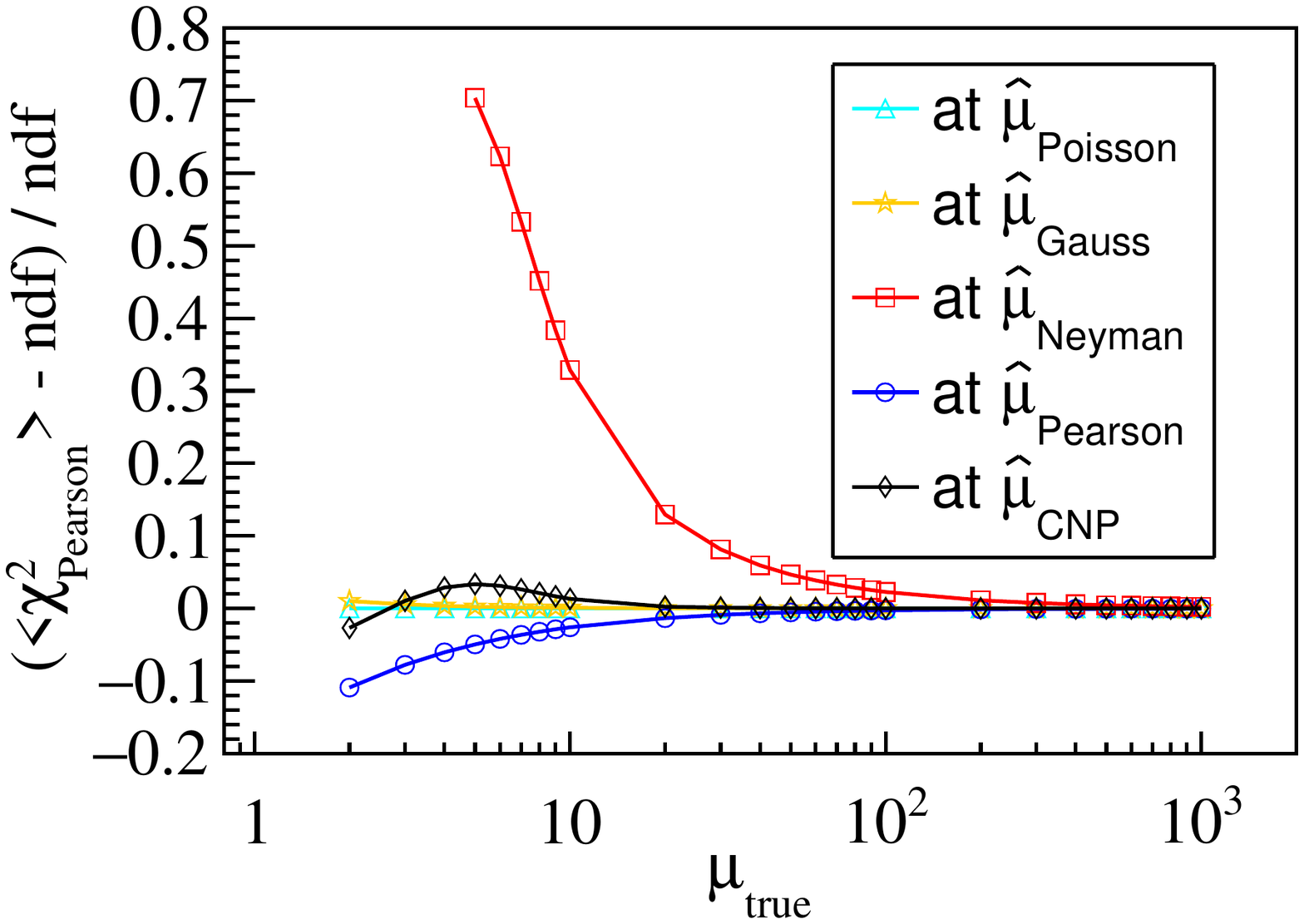}    
        \caption{(Left) Similar to the right panel of Fig.~\ref{Fig::goodness-of-fit} but evaluate the $\chi^2$ at each test statistic's estimator $\hat\mu$. Correspondingly, the resulting $\chi^2$ value is at its minimum $\chi^2_\textrm{min}$. (Right) Similar to the left panel, but use Pearson's chi-square evaluated at a $\hat\mu$ obtained from a different test statistic. Each point is obtained with 10 million toy experiments in the $n=10$ setting. The number of degrees of freedom in all cases is $n-1$ (ndf = 9).}
        \label{Fig::goodness-of-fit-best}
    \end{center}
\end{figure}

In practice, $\mu_\mathrm{true}$ is unknown and experiments often report $\chi^2_{\textrm{min}}$ (evaluated at $\hat\mu$) as a metric for the goodness-of-fit test. The left panel of Fig.~\ref{Fig::goodness-of-fit-best} shows the results of this test for the same setting of $n=10$ as in Fig.~\ref{Fig::goodness-of-fit}. Note that when  $\chi^2$ is evaluated at $\hat\mu$, the number of degrees of freedom is decreased by one (ndf = 9). We see that all five test statistics yield poor results in this goodness-of-fit metric when $\mu_\mathrm{true}$ is less than $\sim$10, indicating large deviations from the chi-square distribution in those cases. On the other hand, inspired by Fig.~\ref{Fig::goodness-of-fit}, we can use $\chi^2_\mathrm{Pearson}$ to perform the goodness-of-fit test, but evaluate it at a $\hat\mu$ obtained from a different test statistic. The right panel of Fig.~\ref{Fig::goodness-of-fit-best} shows the results. We see that when $\chi^2_\mathrm{Pearson}$ is evaluated at a less-biased estimator $\hat\mu$, (e.g.~$\hat\mu_\textrm{Poisson}$, $\hat\mu_\textrm{Gauss}$, or $\hat\mu_\textrm{CNP}$), it results in a better metric for the goodness-of-fit test, which confirms our recommendation in Sec.~\ref{sec:goodness-of-fit}.

\begin{figure}[!htbp]
    \begin{center}
        \includegraphics[angle=0, width=7.5cm] {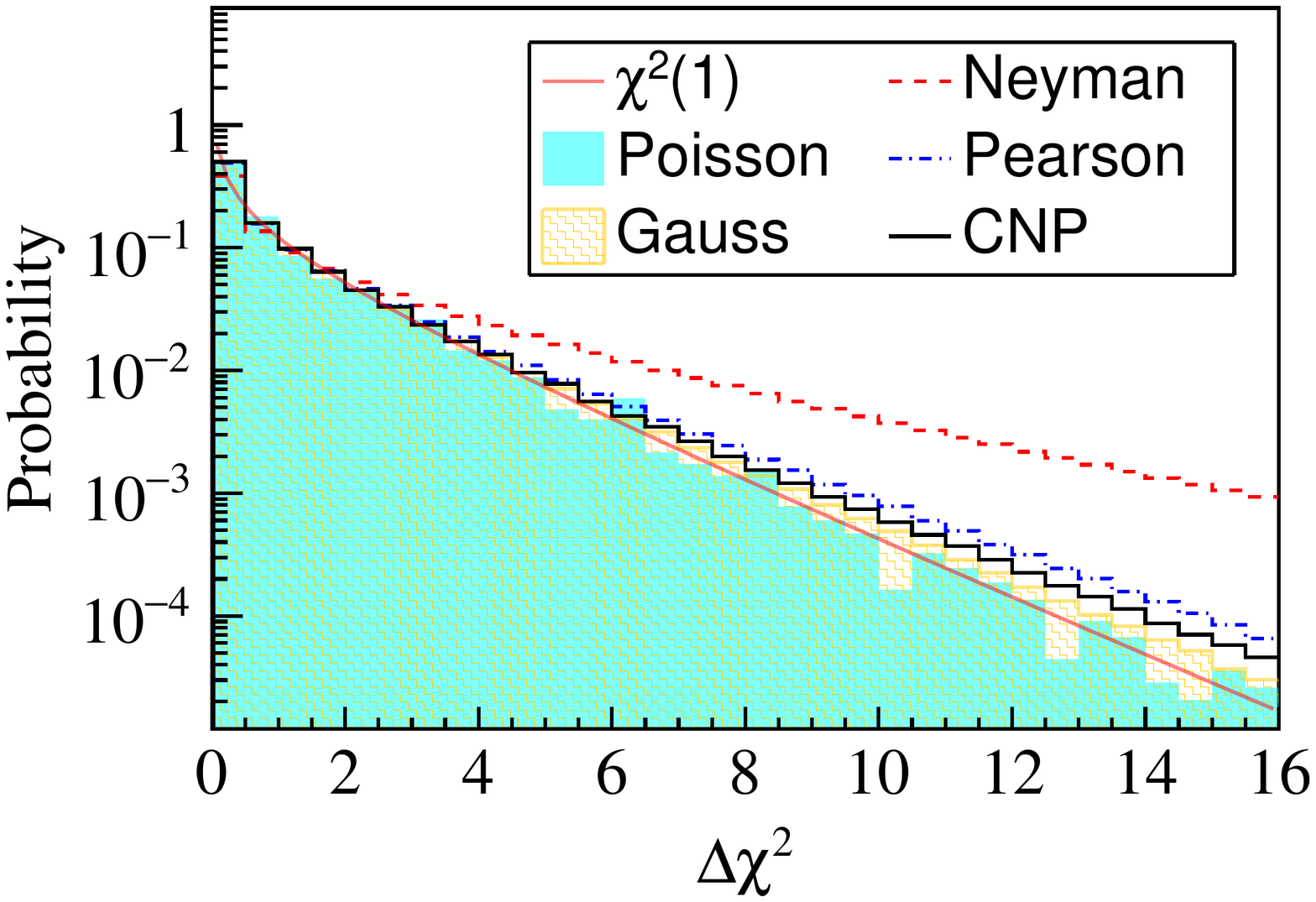}	
        \caption{The distributions of $\Delta\chi^2 $ for the five test statistics: $\chi^2_\mathrm{Poisson}$, 
            $\chi^2_\mathrm{Gauss}$, $\chi^2_\mathrm{Neyman}$, $\chi^2_\mathrm{Pearson}$, and $\chi^2_\mathrm{CNP}$ with $n=10$ and 
            $\mu_\mathrm{true}=15$ using 10 million toy experiments. The ideal 
            chi-square distribution with one degrees of freedom, $\chi^2(1)$, is also shown for comparison.}
        \label{Fig::delta-chi2}
    \end{center}
\end{figure}

To compare the performance on the interval estimation, Fig.~\ref{Fig::delta-chi2} shows 
the $\Delta\chi^2$ distribution in the $n=10$ and $\mu_\textrm{true}=15$ setting 
with 10 million toy experiments, where $\Delta\chi^2 = \chi^2(\mu=\mu_\textrm{true}) 
- \chi^2(\mu=\hat\mu)$. As discussed in Sec.~\ref{sec:CI}, when the conditions of the Wilks' theorem~\cite{Wilks} are met, 
it is expected that $\Delta\chi^2$ in this example follows the chi-square distribution 
with one degree of freedom. However, except for $\chi^2_\mathrm{Poisson}$, the other four test statistics all clearly deviate from the ideal $\chi^2(1)$ distribution leading to 
improper coverages when using the $\Delta\chi^2=1$ rule to set the 68\% confidence intervals.  Therefore, we follow the Feldman-Cousins approach~\cite{Feldman:1997qc} to 
construct the 68\% confidence interval instead. First, a scan of $\mu$ values is performed. 
Setting each test $\mu$ as the true value, many toy experiments are generated to obtain 
its $\Delta\chi^2$ distribution. Then, from each $\Delta\chi^2$ distribution, a 
critical $\Delta\chi^2_c(68\%)$ value can be determined such that below it the 
distribution contains 68\% of the toy experiments. For example, given the distributions 
shown in Fig.~\ref{Fig::delta-chi2}, the critical $\Delta\chi^2_c(68\%)$ values for 
$\chi^2_\mathrm{Neyman}$, $\chi^2_\mathrm{Pearson}$, $\chi^2_\mathrm{Gauss}$, and 
$\chi^2_\mathrm{CNP}$ are larger than one, which is the result of their biases in 
$\hat{\mu}$. Finally, returning to the original toy experiments with the $\mu_\textrm{true}=15$ 
setting, for each toy experiment we can set its confidence interval by comparing 
its $\Delta\chi^2$ value with the critical $\Delta\chi^2_c$ value at 
each test $\mu$ value. The 68\% confidence interval is constructed to 
contain all the test $\mu$ values that have $\Delta\chi^2 < \Delta\chi^2_c(68\%)$.
For each of the 10 million toy experiments, this procedure is repeated to obtain its 68\% confidence 
interval. The reported lower limit $\hat\mu^{-\sigma}_{1/2}$ and upper limit $\hat\mu^{+\sigma}_{1/2}$ of the 68\% confidence interval are the median values 
over all toy experiments and tabulated in Table.~\ref{Tab:SimpleToy}. As shown, $\chi^2_\mathrm{CNP}$ 
and $\chi^2_\mathrm{Gauss}$ have similar (average) interval sizes, both larger than that of $\chi^2_\mathrm{Poisson}$
but quite smaller than those of $\chi^2_\mathrm{Pearson}$ and $\chi^2_\mathrm{Neyman}$. 
There are two reasons causing the larger interval size of $\chi^2_\mathrm{Pearson}$ and $\chi^2_\mathrm{Neyman}$. 
First, $\hat\mu_\mathrm{Neyman}$ has a notably larger variance as shown in Fig.~\ref{Fig::SimpleToy}. 
Second, since $\mu_\textrm{true}$ is always contained in the ensemble median of confidence intervals (but not necessarily near the center) by construction\footnote{In a frequentist definition of the 68\% confidence interval (C.I.), if one performs a large number of similar experiments, the interval would contain $\mu_{\mathrm{true}}$ in 68\% of the cases. This means the lower limit of the 68\% C.I.~would be lower than $\mu_{\mathrm{true}}$ in at least 68\% of the experiments, therefore the median of the lower limit of the 68\% C.I., $\hat\mu^{-\sigma}_{1/2}$, is always lower than $\mu_{\mathrm{true}}$. Similarly, the median of the upper limit of the 68\% C.I., $\hat\mu^{+\sigma}_{1/2}$, is always higher than $\mu_{\mathrm{true}}$.}, the larger biases 
of $\hat\mu_\mathrm{Pearson}$ and $\hat\mu_\mathrm{Neyman}$ also contribute to their larger 
interval sizes. 

\begin{table}[!htbp]
	\caption{Comparison of the median 68\% confidence intervals for the five test statistics: 
$\chi^2_\mathrm{Poisson}$, $\chi^2_\mathrm{Gauss}$, $\chi^2_\mathrm{Neyman}$, $\chi^2_\mathrm{Pearson}$,
and $\chi^2_\mathrm{CNP}$. 10 million toy experiments are generated with the $n=10$ and $\mu_\textrm{true}=15$ 
setting. For each toy experiment, a 68\% confidence interval is obtained using the Feldman-Cousins approach. 
The reported lower limit $\hat\mu^{-\sigma}_{1/2}$ and upper limit $\hat\mu^{+\sigma}_{1/2}$ of 
the 68\% confidence interval are the median values over all toy experiments. }
	\label{Tab:SimpleToy}
	\begin{center}
		\begin{tabular}[c]{|c|c|c|}\hline
			& median 68\% confidence interval & interval size \\ 
			& $\left(\hat\mu^{-\sigma}_{1/2}, \, \hat\mu^{+\sigma}_{1/2}\right)$ & $\hat\mu^{+\sigma}_{1/2} - \hat\mu^{-\sigma}_{1/2}$ \\ \hline
			$\chi^2_\textrm{Poisson}$ & (13.839, 16.226) & 2.387 \\ \hline
			$\chi^2_\textrm{Gauss}$ & (13.744, 16.221) & 2.478 \\ \hline
			$\chi^2_\textrm{Neyman}$ & (12.236, 15.706) & 3.471 \\ \hline
			$\chi^2_\textrm{Pearson}$ & (14.153, 16.800) & 2.647 \\ \hline
			$\chi^2_\textrm{CNP}$ & (13.745, 16.196) & 2.451 \\ \hline
		\end{tabular}
	\end{center}
\end{table}

Next, we show two more examples with increasing complexity inspired by 
real experiments. Since $\chi^2_\textrm{Gauss}$ generally have a similar performance as $\chi^2_\textrm{CNP}$ and can also benefit from 
the covariance matrix formalism, we restrict our comparisons of $\chi^2_\textrm{CNP}$ to 
$\chi^2_{\mathrm{Poisson}}$, $\chi^2_{\mathrm{Neyman}}$, and $\chi^2_{\mathrm{Pearson}}$.
The following study will focus on the bias of the point estimation of model parameters, since the performance on the goodness-of-fit test and the interval estimation is similar to the first example. 

\subsection{Example 2: fitting multi-detector histograms}
\label{sec:example_2}

In this section, we introduce a more realistic example, which is inspired by the PROSPECT reactor neutrino
experiment~\cite{Ashenfelter:2018zdm} searching for a light sterile neutrino~\cite{Ashenfelter:2018iov}.
One of the unique features of PROSPECT is that the detector consists of many segmented sub-detectors,
and the number of events in each sub-detector is not high ($\sim$few hundreds).
Since each sub-detector has a different baseline to the reactor, it is desirable to treat each sub-detector separately
in the spectrum fitter to increase the physics sensitivity to the energy- and baseline-dependent
oscillation effect caused by a hypothetical light sterile neutrino.

In our toy example experiment, we assume 100 sub-detectors, each measures a common energy spectrum with 16 energy bins.
The expected spectrum is assumed to be flat with an unknown normalization bias factor $\epsilon$ to be determined\footnote{\ref{sec:appendixE} shows an example where the shape of the histogram is also a model parameter.}.
In the $i$th bin of the $d$th detector, $\mu_d^i$ signal events and $b_d^i$ background events are expected,
and $M_d^i$ total events are measured. The background shape is also assumed to be flat and the expected background $b_d^i$ is assumed to be half of the expected signal $\mu_d^i$ in size.
The experiment also measured $B_d^i$ background events in a signal-off period, which provided an external constraint on the background. For simplicity we assume the length of the signal-off period  is the same as the signal-on period.
We consider one systematic uncertainty, the relative normalization uncertainty $\epsilon_d$ among detectors,
and assume it to be constrained to 2\%. Therefore, in this example, there is one model parameter $\epsilon$,
and 1700 nuisance parameters ($b_d^i$, $\epsilon_d$) to be estimated.

The Poisson-likelihood chi-square function for this toy experiment can be written as:
\begin{equation}
  \begin{aligned}
   \label{eq:chi2_Poisson_full_syst_bkg}
   \chi^2_\mathrm{Poisson} = & 2\sum_{d=1}^{100}\sum_{i=1}^{16}\left(\mu_d^i(1+\epsilon+\epsilon_d)+b_d^i-M_d^i+M_d^i\mathrm{ln}\frac{M_d^i}{\mu_d^i(1+\epsilon+\epsilon_d)+b_d^i}\right)\\
   & + 2\sum_{d=1}^{100}\sum_{i=1}^{16}\left(b_d^i-B_d^i+B_d^i\mathrm{ln}\frac{B_d^i}{b_d^i}\right)+ \sum_{d=1}^{100}\left(\frac{ \epsilon_d}{0.02}\right)^2,
  \end{aligned}
\end{equation}
and for the CNP chi-square:
\begin{equation} \label{eq:chi2_CNP_full_syst_bkg}
   \chi^2_\mathrm{CNP} =  \sum_{d=1}^{100}\sum_{i=1}^{16}\frac{\left(\mu_d^i(1+\epsilon+\epsilon_d)+b_d^i-M_d^i \right)^2}{3/(\frac{1}{M_d^i}+\frac{2}{\mu_d^i(1+\epsilon+\epsilon_d)+b_d^i})} 
    + \sum_{d=1}^{100}\sum_{i=1}^{16} \frac{\left(b_d^i-B_d^i \right)^2}{3/(\frac{1}{B_d^i}+\frac{2}{b_d^i})}+ \sum_{d=1}^{100}\left(\frac{ \epsilon_d}{0.02}\right)^2.
\end{equation}
The $\chi^2_{\mathrm{Neyman}}$ and $\chi^2_{\mathrm{Pearson}}$ can be constructed similarly by changing the denominators of
the first and the second terms in Eq.~\eqref{eq:chi2_CNP_full_syst_bkg}. 

Minimizing the above chi-square functions involves finding the best-fit values of
the 1700 nuisance parameters, which could cause instabilities of the fitter. To reduce the number of nuisance parameters,
we can find their best-fit values by solving the corresponding differential equations, e.g.~$\partial \chi^2/\partial b_d^i = 0$.
In this simple example, since the nuisance parameters are independent of each other, this equation is linear for $\chi^2_{\mathrm{Neyman}}$, quadratic for $\chi^2_{\mathrm{Poisson}}$,
quartic for $\chi^2_{\mathrm{Pearson}}$, and quintic for $\chi^2_{\mathrm{CNP}}$. The solutions to these equations can be found
either analytically ($\leq4^{\mathrm{th}}$ order) or numerically ($>4^{\mathrm{th}}$ order).

\begin{figure}[!htbp]
	\begin{center}
		\includegraphics[angle=0, width=7.5cm] {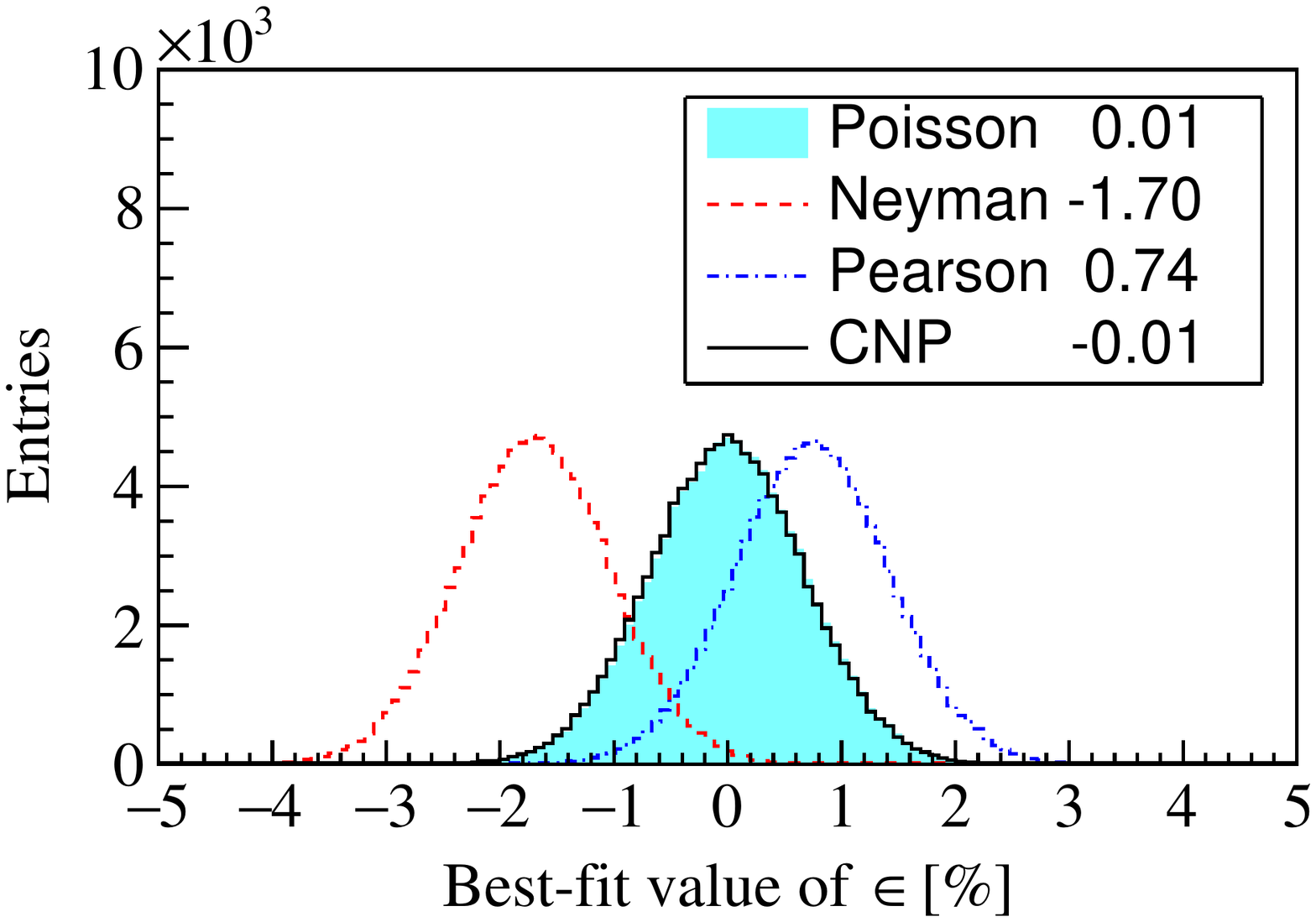}	
		\caption{Distributions of best-fit values of the normalization bias factor $\epsilon$
			using  $\chi^2_\mathrm{Poisson}$, $\chi^2_\mathrm{Neyman}$, $\chi^2_\mathrm{Pearson}$, and $\chi^2_\mathrm{CNP}$.
			One hundred thousand toy experiments are simulated. Each experiment has 100 detectors and 16 energy bins.
			The nominal signal and background in each bin are assumed to be 30 and 15, respectively. The numbers in the legend show the mean of each distribution.}
		\label{Fig::SecondToy}
	\end{center}
\end{figure}

One hundred thousand toy experiments are simulated assuming the nominal signal $\mu_d^i = 30$ and background $b_d^i = 15$
in each bin. The normalization bias factor $\epsilon$ is fitted for each experiment,
where the true value of $\epsilon$ is set to zero.
The results of using $\chi^2_\mathrm{Poisson}$, $\chi^2_\mathrm{Neyman}$, $\chi^2_\mathrm{Pearson}$ and $\chi^2_\mathrm{CNP}$
are shown in Fig.~\ref{Fig::SecondToy}.
Despite being small, the bias of $\chi^2_\mathrm{Poisson}$ is non-zero. This is caused by the introduction of penalty terms
in Eq.~\eqref{eq:chi2_Poisson_full_syst_bkg} (see \ref{sec:appendixC} for an explanation).
One can see that the bias of $\chi^2_\mathrm{CNP}$ is again much smaller than
those of $\chi^2_\mathrm{Neyman}$ and $\chi^2_\mathrm{Pearson}$, representing a much better approximation to $\chi^2_\mathrm{Poisson}$.

\subsection{Example 3: covariance matrix implementation}
\label{sec:example_3}

In many physics experiments, covariance matrix is used to model complicated systematic uncertainties,
where either direct nuisance parameter implementation is difficult, or there are too many nuisance parameters to minimize.
In this section, we show how the $\chi^2_\mathrm{CNP}$ can be implemented in a covariance matrix format.

We introduce a slight complication to the previous example so that the analytic or numerical methods to find best-fit values are
prohibitively difficult in the minimization.
In this example, we assume the detector response changed between the signal-on and the signal-off period,
and in order to interpolate the expected background in the signal-off period $b_d^i$ to the signal-on period,
a transfer matrix $R$ is needed such that $(b_d^i)_{\mathrm{on}}=\sum_{j}R_d^{ij}b_d^j$.
For simplicity, 10 sub-detectors are used in this example, and the transfer matrix $R$ does a simple smearing
in energy bins such that for each detector $R^{ij}_d=0.5$ when $i=j$, $R^{ij}_d=0.25$ when $i=j\pm1$, and $R^{ij}_d=0$ everywhere else.
The $\chi^2_\mathrm{CNP}$ in this example becomes:
\begin{equation}\label{eq:chi2_CNP_full_third_syst_bkg}
  \begin{aligned}
    \chi^2_\mathrm{CNP} =  &\sum_{d=1}^{10}\sum_{i=1}^{16}\frac{\left(\mu_d^i(1+\epsilon+\epsilon_d)+\sum_{j}R_d^{ij}b_d^j-M_d^i \right)^2}{3/(\frac{1}{M_d^i}+\frac{2}{\mu_d^i(1+\epsilon)+\sum_{j}R_d^{ij}b_d^j})} 
    + \sum_{d=1}^{10}\sum_{i=1}^{16} \frac{\left(b_d^i-B_d^i \right)^2}{3/(\frac{1}{B_d^i}+\frac{2}{b_d^i})}\\
    &+ \sum_{d=1}^{10}\left(\frac{ \epsilon_d}{0.02}\right)^2.
  \end{aligned}
\end{equation}
In this case, solving for the nuisance parameters through $\partial \chi^2/\partial b_d^i = 0$ would lead to a set of quintic equations,
which is difficult to solve either analytically or numerically. Following Sec.~\ref{sec:cnp_construction}, the covariance matrix format of Eq.~\eqref{eq:chi2_CNP_full_third_syst_bkg} is:
\begin{equation}
\label{eq:cov_chi2_CNP_full_third}
\begin{aligned}
\left(\chi^2_{\mathrm{CNP}}\right)_{\mathrm{cov}} = 
& \left( {\bm{\mu}}(1+\epsilon) + R\cdot {\bm b} - {\bm M} \right)^T \cdot \left(V^{\mathrm{stat}}_{\mathrm{CNP}}+V^{\mathrm{syst}}\right)^{-1} \cdot \left( { \bm{\mu}}(1+\epsilon) + R\cdot {\bm b} - {\bm M} \right) \\
& +
\left( {\bm{b}} - {\bm B} \right)^T \cdot \left(V^{\mathrm{bkg}}_{\mathrm{CNP}}\right)^{-1} \cdot \left( {\bm{b}} - {\bm B} \right),
\end{aligned}
\end{equation}
where $M_d^i$, $\mu_d^i$, $b_d^i$ and $B_d^i$ are ordered into a single 160-element
vector $\bm M$, $\bm \mu$, $\bm b$, $\bm B$, respectively.
$V^{\mathrm{stat}}_{\mathrm{CNP}}$ is the covariance matrix corresponding to the statistical uncertainty,
which is diagonal with its elements being the corresponding values in the denominator of the first
term of Eq.~\eqref{eq:chi2_CNP_full_third_syst_bkg}.
Similarly , $V^{\mathrm{bkg}}_{\mathrm{CNP}}$ is the covariance matrix corresponding to
the background statistical uncertainty with the diagonal elements defined by the denominator of the second term
in Eq~\eqref{eq:chi2_CNP_full_third_syst_bkg}.
$V^{\mathrm{syst}}$ is the covariance matrix corresponding to the systematic uncertainty $\epsilon_d$,
which can be calculated either analytically or from toy Monte Carlo simulations
by randomly fluctuating the number of events according to $\epsilon_d$ and its constraint.

\begin{figure}[!htbp]
	\begin{center}
		\includegraphics[angle=0, width=6.8cm] {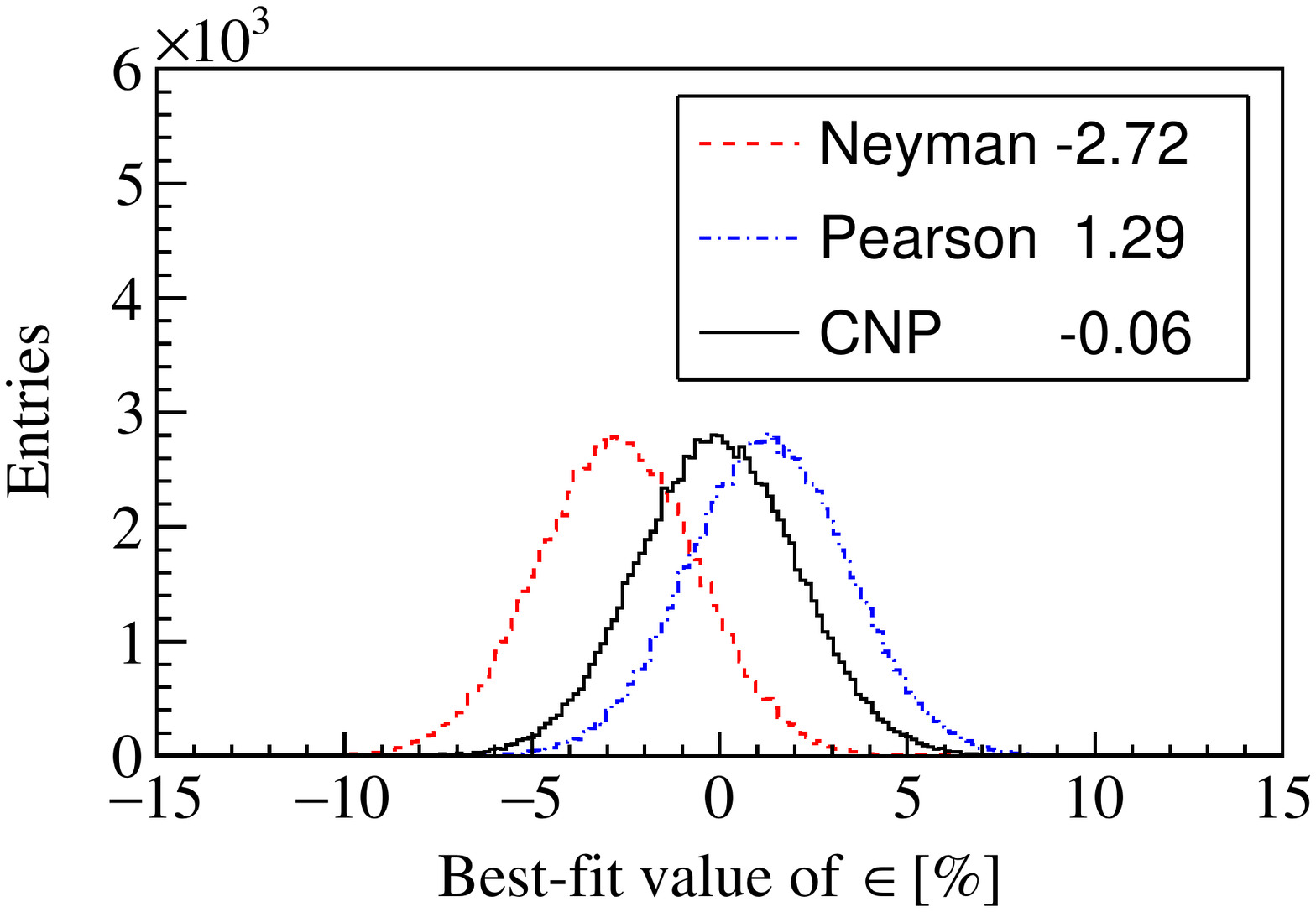}	
		\includegraphics[angle=0, width=6.8cm] {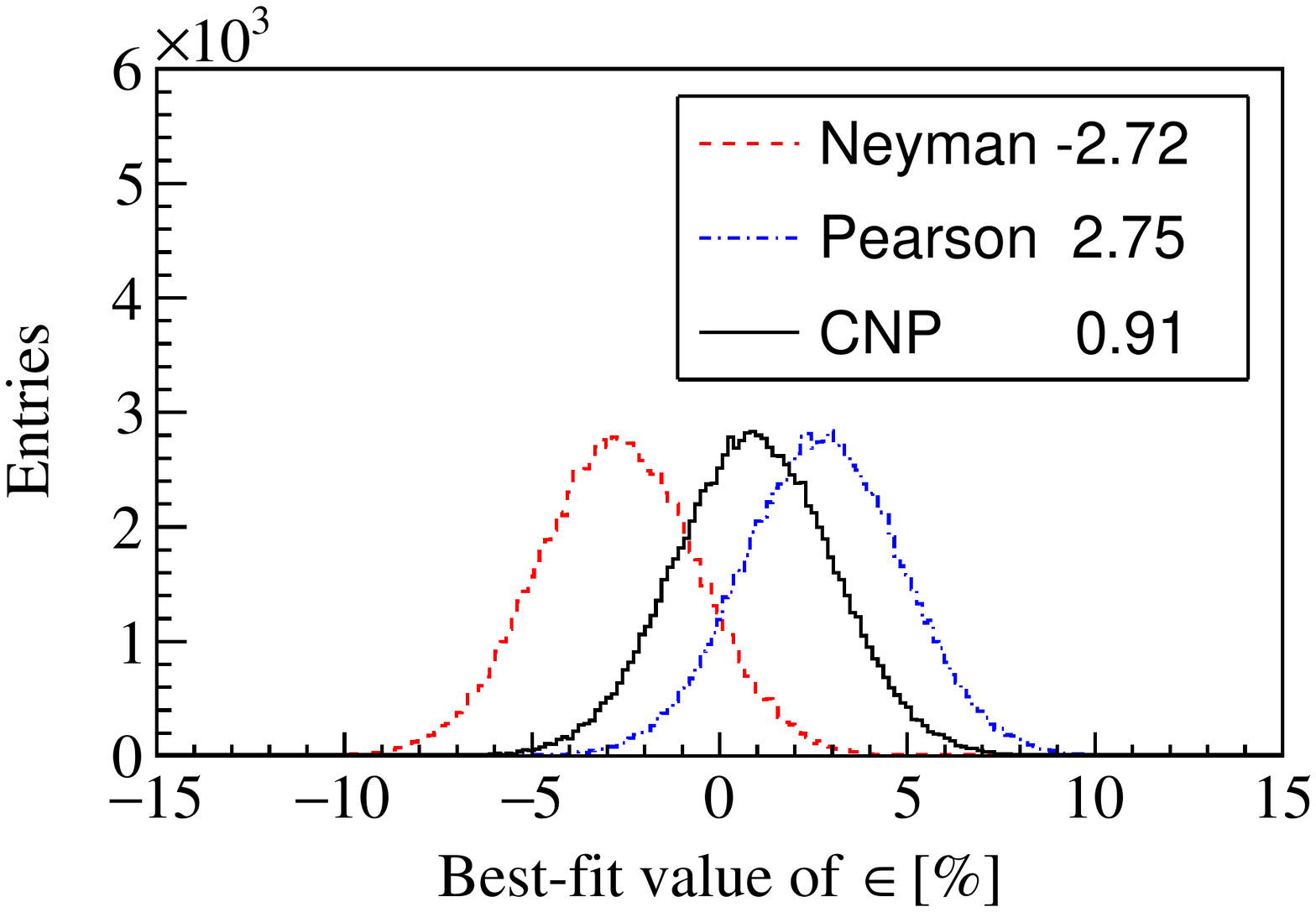}    
		\caption{(Left) Distributions of best-fit values of normalization factor $\epsilon$ from
                  $(\chi^2_\mathrm{Neyman})_\mathrm{cov}$, $(\chi^2_\mathrm{Pearson})_\mathrm{cov}$ and $(\chi^2_\mathrm{CNP})_\mathrm{cov}$ in the third example, simulated using one hundred thousand toy experiments with 10 sub-detectors and $\mu_d^i=30$ and $b_d^i=15$. The numbers in the legend show the mean of each distribution.
                  (Right) Similar to the left plot but after further approximation to absorb the background term into the covariance matrix as in Eq.~\ref{eq:chi2_CNP_full_third_approx} and Eq~\eqref{eq:cov_chi2_CNP_full_third_approx}.}
		\label{Fig::ResutlThirdExample}
	\end{center}
\end{figure}

Following the same procedure, covariance matrix formats can be constructed for $\chi^2_\mathrm{Pearson}$
and $\chi^2_\mathrm{Neyman}$ by replacing the statistical uncertainty terms in the covariance matrix
in Eq.~\eqref{eq:cov_chi2_CNP_full_third}, $V^{\mathrm{stat}}_{\mathrm{CNP}}$ and $V^{\mathrm{bkg}}_{\mathrm{CNP}}$, to their corresponding values in $\chi^2_\mathrm{Pearson}$ and $\chi^2_\mathrm{Neyman}$.
We note that there is no equivalent covariance matrix format for the Poisson-likelihood chi-square.
One hundred thousand toy experiments are simulated assuming the nominal signal $\mu_d^i = 30$ and background $b_d^i=15$ in each bin.
The normalization bias factor $\epsilon$ is fitted for each experiment,
where the true value of $\epsilon$ was set to zero.
The results are shown in the left panel of Fig.~\ref{Fig::ResutlThirdExample}.
We see that in the covariance format, the bias of $(\chi^2_\mathrm{CNP})_\mathrm{cov}$ is again more than
an order of magnitude smaller than those of $(\chi^2_\mathrm{Neyman})_\mathrm{cov}$ and $(\chi^2_\mathrm{Pearson})_\mathrm{cov}$.

We emphasize that in the $(\chi^2_\mathrm{CNP})_\mathrm{cov}$ defined in Eq.~\eqref{eq:cov_chi2_CNP_full_third},
both the free parameter $\epsilon$ and the nuisance parameters $b^d_i$ need to be minimized.
This is due to the nature of the Poisson statistical uncertainty of the background, and how it is treated in the CNP chi-square.
It is tempting to further reduce the number of nuisance parameters by absorbing them into a fixed covariance matrix.
In order to do so, we need to approximate the expected $b^d_i$ with their measured value $B^d_i$.
In this case, Eq.~\eqref{eq:chi2_CNP_full_third_syst_bkg} and \eqref{eq:cov_chi2_CNP_full_third} are replaced by
\begin{equation}\label{eq:chi2_CNP_full_third_approx}
\begin{aligned}
\chi^{'2}_\mathrm{CNP} = &\sum_{d=1}^{10}\sum_{i=1}^{16}\frac{\left(\mu_d^i(1+\epsilon+\epsilon_d)+\sum_{j}R_d^{ij}b_d^j-M_d^i \right)^2}{3/(\frac{1}{M_d^i}+\frac{2}{\mu_d^i(1+\epsilon)+\sum_{j}R_d^{ij}B_d^j})} 
+ \sum_{d=1}^{10}\sum_{i=1}^{16} \frac{\left(b_d^i-B_d^i \right)^2}{B_d^i}\\
&+ \sum_{d=1}^{10}\left(\frac{ \epsilon_d}{0.02}\right)^2
\end{aligned}
\end{equation}
and
\begin{equation}\label{eq:cov_chi2_CNP_full_third_approx}
\begin{aligned}
\left(\chi^{'2}_{\mathrm{CNP}}\right)_{\mathrm{cov}} 
= &\left( {\bm{\mu}}(1+\epsilon) + R\cdot {\bm B} - {\bm M} \right)^T \cdot \left(V^{'\mathrm{stat}}_{\mathrm{CNP}}+V^{'\mathrm{bkg}}+V^{'\mathrm{syst}}\right)^{-1}\\
&\cdot \left( { \bm{\mu}}(1+\epsilon) + R\cdot {\bm B} - {\bm M} \right),
\end{aligned}
\end{equation}
where the nuisance parameters $b_d^i$ are absorbed into $V^{'\mathrm{bkg}}$. 
After these approximations, in $\left(\chi^{'2}_{\mathrm{CNP}}\right)_{\mathrm{cov}}$,
only one free parameter $\epsilon$ instead of 161 fitting parameters in Eq.~\eqref{eq:cov_chi2_CNP_full_third} needs
to be minimized and the computational cost is largely reduced.
Similar approximations can be used for $\left(\chi^{'2}_{\mathrm{Pearson}}\right)_{\mathrm{cov}}$ and
the fitting results are shown in right panel of Fig.~\ref{Fig::ResutlThirdExample}.
We see that although the approximation leads to a much reduced number of fitting parameters,
the bias of the normalization factor $\epsilon$ becomes significantly larger, in particular for the CNP-chi-square. It is therefore crucial to indicate clearly how the $\chi^2$ is defined,
and what approximations are implied in the construction of the covariance matrix when reporting results.

\section{Discussions}\label{sec:discussion}

Through examples in the previous section, we have compared various chi-square construction methods and different minimization strategies. In the following, we provide some recommendations on when to use them in the data analysis of counting experiments:
\begin{itemize}
\item When the computational cost is not a concern (e.g.~number of nuisance parameters is small), a direct minimization of the Poisson-likelihood chi-square (with nuisance parameters implementing through pull terms) should be used. 
\item When the computational cost of a direct minimization is high, one should first look for analytic or numerical solutions, which can effectively reduce the number of nuisance parameters without making any approximations. 
For example, the number of nuisance parameters of the Poisson-likelihood chi-square in the example described in Sec.~\ref{sec:example_2} can be reduced by solving a set of independent quadratic equations.
\item When analytic or numerical solutions are not available, approximations may become necessary to reduce the computational cost. In this case, the covariance matrix formalism is a common tool in reducing the number of nuisance parameters. 
However, before approximating the Poisson-likelihood chi-square by Neyman's, Pearson's, Gauss-likelihood, or CNP chi-squares, one can examine if it is sufficient to apply covariance matrix only to the pull terms of the systematic uncertainties. 
For example, the rate plus shape oscillation fit described in Ref.~\cite{An:2013zwz} used a covariance matrix in the pull term for reactor-related uncertainties. In this approach, the statistical part of 
the chi-square function can still use the Poisson-likelihood format. 
\item When the Poisson-likelihood chi-square has to be replaced, the iterative approach with the weighted least-squares as described in Ref.~\cite{stat1, stat2, Dembinski:2018ihc} can be an option to eliminate the bias in the estimator. 
An alternative approach is the CNP or the Gauss-likelihood chi-square, which both lead to a much reduced bias in estimating model parameters than using either Neyman's or Pearson's chi-square. 
As shown in Fig.~\ref{Fig::Bias_vs_N} of Sec.~\ref{sec:example_1}, the CNP or the Gauss-likelihood chi-square could be the better choice of test statistics depending on the number of measurements.
In addition, the improved confidence intervals (smaller in size or with more proper coverage) are often accompanied with the reduced bias as discussed in Sec.~\ref{sec:CI} and shown in Sec.~\ref{sec:example_1}.
Similarly, analytic or numerical solutions should be explored before applying a covariance matrix approach, since additional approximations are necessary in the later case. 
As shown in Sec.~\ref{sec:cnp_construction}, the derivation of covariance matrix formula assumes i) the variance describing statistical fluctuations has to be independent of any nuisance parameters, and ii) the predicted counts only have a linear dependence on the nuisance parameters. 
For example, the approximation made in the right panel of Fig.~\ref{Fig::ResutlThirdExample} leads to a significant bias.
\end{itemize} 

We emphasize that since there are many different ways to make approximations in defining the chi-square test statistics, it is extremely important for experiments to clearly report how their test statistics are constructed. 

In summary, we proposed a linear combination of Neyman's and Pearson's chi-squares, $\chi^2_{\mathrm{CNP}}$, as an improved approximation to the widely-used Poisson-likelihood chi-square in counting experiments. With three examples, we show that the bias in parameter estimation from using CNP chi-square is much smaller than those using the Neyman's or 
Pearson's chi-square alone. In occasions where the computational cost of using Poisson-likelihood chi-square is high, the CNP chi-square with its covariance matrix format provides a good alternative. 

\section*{Acknowledgments}
We thank Maxim Gonchar and Mike Shaevitz for suggesting the comparison of the 
CNP chi-square with the Gauss-likelihood chi-square. This work is supported by 
the U.S. Department of Energy, Office of Science, Office of High Energy 
Physics, and Early Career Research Program under contract number DE-SC0012704.


\appendix
\section{Treatment of bins with zero observed events} \label{sec:appendixA}
Experiments can often have bins with zero counts when the expected signal is small.
In this case, the Neyman's chi-square definition, Eq.~\eqref{eq:Neyman-chi2}, breaks down since the measured number of events is
in the denominator, so are the CNP and Gauss-likelihood chi-square definitions.
Practical approximations are often made in experiments by either ignoring bins with zero observation, or assign the statistical uncertainty as 1 for 
zero-count bins (e.g.~the ``modified Neyman's chi-square''~\cite{comp_teststat}).
Here we adopt the Poisson-likelihood chi-square definition for zero-count bins:
\begin{equation} \label{eq:chi2_zeroBin}
\left(\chi^2_i\right)_{M_i=0}
=2\left(\mu_i({\bf\bm\theta})-M_i+M_i\ln\frac{M_i}{\mu_i({\bf\bm\theta})}\right)_{M_i=0}
=2\mu_i({\bf\bm\theta}). 
\end{equation}
Eq.~\eqref{eq:chi2_zeroBin} can be re-written in a weighted least-squares format: 
\begin{equation}\label{eq:least-square_zeroBin} 
\left(\chi^2_i\right)_{M_i=0}
=2\mu_i({\bf\bm\theta})=\frac{(\mu_i({\bf\bm\theta})-M_i)^2}{\mu_i({\bf\bm\theta})/2}.
\end{equation}
Compared with the Pearson's chi-square, we see that the variance is half of $\chi^2_{\mathrm{Pearson}}$ for zero-count bins.
The covariance matrix element corresponding to a zero-count bin follows:
\begin{equation}
\label{eq:cov_zeroBin}
\left(V^{\mathrm{stat}}({\bf\bm\theta})_{ij}\right)_{M_i=0} = \frac{\mu_i({\bf\bm\theta})}{2} \delta_{ij}.
\end{equation}
In this paper, we use Eq.~\eqref{eq:chi2_zeroBin} and ~\eqref{eq:cov_zeroBin} in all occasions when zero-count bins are encountered.

\section{Bias of estimator $\hat{\mu}_{\mathrm{Neyman}}$ and $\hat{\mu}_{\mathrm{Pearson}}$ versus number of measurements} \label{sec:appendixB}
Here we prove that the bias of $\hat{\mu}_{\mathrm{Neyman}}$ and $\hat{\mu}_{\mathrm{Pearson}}$
increases as the number of measurements $n$ increases, as shown in Fig.~\ref{Fig::Bias_vs_N}.
Making use of the relations
\begin{equation}\label{eq:expectation_relations}
\mathrm{Var}(x) = E(x^2) - E^2(x), \quad
E\left(\frac{1}{x}\right) \approx \frac{1}{E(x)} + \frac{\mathrm{Var}(x)}{E^3(x)},
\end{equation}
for $\hat{\mu}_{\mathrm{Neyman}}$ we have:
\begin{equation}
E\left(\frac{1}{\hat{\mu}_{\mathrm{Neyman}}}\right)  =  E\left(\frac{\sum_{i=1}^{n}\frac{1}{M_i}}{n}\right) = E\left(\frac{1}{M_i}\right) \approx \frac{1}{E(M_i)}+\frac{\mathrm{Var}(M_i)}{(E(M_i))^3} = \frac{1}{\mu} + \frac{1}{\mu^2}, 
\end{equation}
where $E(M_i) = \mathrm{Var}(M_i) = \mu$ since $M_i$ follows a Poisson distribution. The expected bias then becomes:
\begin{equation}\label{eq:Neymanlimit}
E(\hat{\mu}_{\mathrm{Neyman}}-\mu) 
= E\left(\frac{1}{\frac{1}{\hat{\mu}_{\mathrm{Neyman}}}}\right)-\mu
\approx -\frac{\mu}{1+\mu} +  \frac{\mathrm{Var}\left(\frac{1}{M_i}\right)/n}{\left(\frac{1}{\mu} + \frac{1}{\mu^2}\right)^3}.
\end{equation}
which deviates further from zero when $n$ increases. The bias approaches -1 
when $n$ and $\mu$ become large.~\footnote{Note that for the dependence on $\mu$, Eq.~\eqref{eq:Neymanlimit} is only asymptotically correct when $n$ and $\mu$ are large due to the approximation made in Eq.~\eqref{eq:expectation_relations}. The actual dependence on $\mu$ when $n\to\infty$ can only be written as an infinite summation (e.g.~$E(\hat\mu_{\mathrm{Neyman}}) = (e^{\mu} -1)/\left(1+\sum_{k=1}^{\infty}\frac{\mu^k}{k(k!)}\right)$). One derivation can be found in Ref~\cite{chi2-gamma}.}

Similarly, for $\hat{\mu}_{\mathrm{Pearson}}$ we have:
\begin{equation}
\begin{aligned}
  E\left( \hat{\mu}_{\mathrm{Pearson}} \right) &= E\left(\sqrt{\frac{\sum_i M_i^2}{n}} \right) = \sqrt{E(M_i^2)-\mathrm{Var}\left( \hat{\mu}_{\mathrm{Pearson}} \right)}\\
  &= \sqrt{\mu^2 + \mu - \mathrm{Var}\left( \hat{\mu}_{\mathrm{Pearson}} \right)},
\end{aligned}
\end{equation}
therefore:
\begin{equation}
E(\hat{\mu}_{\mathrm{Pearson}}-\mu) = \mu\left(\sqrt{1 + \frac{1}{\mu}-\frac{\mathrm{Var}\left( \hat{\mu}_{\mathrm{Pearson}} \right)}{\mu^2}} - 1\right),
\end{equation}
which also becomes larger at larger $n$, since the variance of 
$\hat{\mu}_{\mathrm{Pearson}}$ becomes smaller at larger $n$. The bias approaches 1/2 when $n$ and $\mu$ become large.

\section{Bias of $\chi^2_\mathrm{Poisson}$ when pull terms are included} \label{sec:appendixC}
In this appendix, we provide an explanation of the non-zero bias
of $\epsilon$ from $\chi^2_\mathrm{Poisson}$ when pull terms are included, for example, in Eq.~\eqref{eq:chi2_Poisson_full_syst_bkg}.
Let us consider a simplified example. One experiment measured $m$ number of events,
which follows Poisson-distribution with the mean value of $\mu$.
There is one systematic uncertainty ($\epsilon$) on the normalization of $\mu$,
which is constrained with standard deviation of $\sigma$.
Following maximum-likelihood principle, the Poisson-likelihood chi-square with the constraint on $\epsilon$ is:
\begin{equation}
\chi^2_\mathrm{Poisson}=2\left(\mu(1+\epsilon)-m+m\cdot\mathrm{ln}\frac{m}{\mu(1+\epsilon)}\right)+\left(\frac{\epsilon}{\sigma}\right)^2.
\end{equation}
The estimator of $\epsilon$ ($\hat{\epsilon}$) can be derived through the minimization of chi-square: $\partial \chi^2_\mathrm{Poisson}/\partial\epsilon=0$:
\begin{equation}
\label{eq:B1_estimator}
\hat{\epsilon} = \frac{1+\mu\sigma^2}{2}\left(-1+\sqrt{1-\frac{4\sigma^2}{(1+\mu\sigma^2)^2}(\mu-m)}\right).
\end{equation}
Defining $x=\frac{4\sigma^2}{(1+\mu\sigma^2)^2}(\mu-m)$ and assuming $|x|\ll1$,
we can perform a Taylor expansion on Eq.~\eqref{eq:B1_estimator}
and obtain:
\begin{equation}
\label{eq:B1_estimator_tayler}
\hat{\epsilon} \approx\frac{1+\mu\sigma^2}{4}\left(-x-\frac{1}{4}x^2-O(x^3)\right).
\end{equation}
Ignoring higher-order terms, the expectation of $\hat{\epsilon}$ is
\begin{equation}
\label{eq:B1_estimator_expectation}
E(\hat{\epsilon}) \approx\frac{1+\mu\sigma^2}{4}\left(-E(x)-\frac{1}{4}E(x^2)\right).
\end{equation}
Given that $E(x)$ is zero and $E(x^2)$ is non-zero, we see that in this example $\hat{\epsilon}$ is a biased estimator.
$\hat{\epsilon}$ only asymptotically becomes unbiased under large statistics~\cite{pdg}. 

\section{Bias and covariance matrix formulas for the Gauss-likelihood chi-square}~\label{sec:appendixD}
In this appendix, we provide formulas on the bias of $\hat\mu_{\mathrm{Gauss}}$ from the Gauss-likelihood chi-square $\chi^2_{\mathrm{Gauss}}$, as well as the covariance 
matrix format of $\chi^2_{\mathrm{Gauss}}$. Given the simple model described in Sec.~\ref{sec:cnp_construction}, 
$\hat{\mu}_{\mathrm{Gauss}}$ can be obtained through the minimization of 
Eq.~\eqref{eq:chi2_Gauss}: $\partial \chi^2_{\mathrm{Gauss}}/\partial \mu = 0$, yielding
\begin{equation}
\hat{\mu}_{\mathrm{Gauss}} = \sqrt{\frac{\sum_{i=1}^n M_i}{n} + \frac{1}{4}} - \frac{1}{2} \,.
\end{equation}

Using the covariance matrix formalism, the likelihood function in 
Eq.~\eqref{eq:Gauss-likelihood} becomes:
\begin{equation}
L_{\mathrm{Gauss}}({\bm\mu}({\bm\theta}); {\bm M}) = \frac{1}{\sqrt{(2\pi)^d\lvert V \rvert}}\cdot  \exp\left[ \frac{1}{2} \left( {\bm M - \bm{\mu}({\bf\bm\theta})} \right)^T \cdot V^{-1} \cdot \left( {\bm M - \bm{\mu}({\bf\bm\theta})} \right) \right],
\end{equation}
where $d$ and $\lvert V \rvert$ are the dimension and determinant of the covariance matrix $V$, respectively. Therefore, we have
\begin{equation}
\chi^2_{\mathrm Gauss} = -2\ln \lambda_{\mathrm Gauss}({\bm \theta}) = \ln \lvert V \rvert  + \left( {\bm M - \bm{\mu}({\bf\bm\theta})} \right)^T \cdot V^{-1} \cdot \left( {\bm M - \bm{\mu}({\bf\bm\theta})} \right) + C \,,
\end{equation}
with $C$ being a model-independent constant, which does not play a role in
estimating the model parameters. 

\section{Improvement on model parameters other than normalization}~\label{sec:appendixE}
\begin{figure}[!htbp]
    \begin{center}
        \includegraphics[angle=0, width=6.8cm] {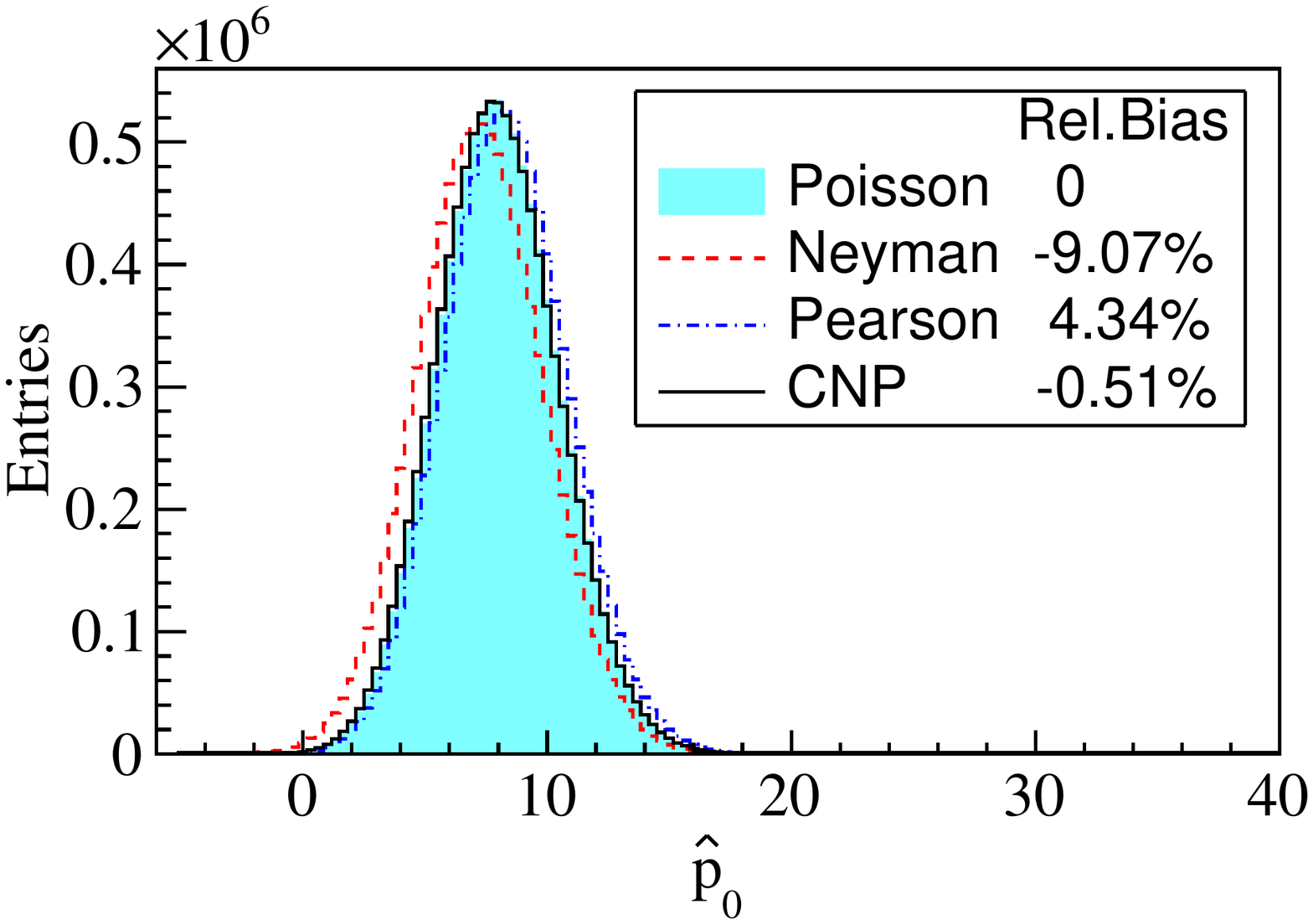}	
        \includegraphics[angle=0, width=6.8cm] {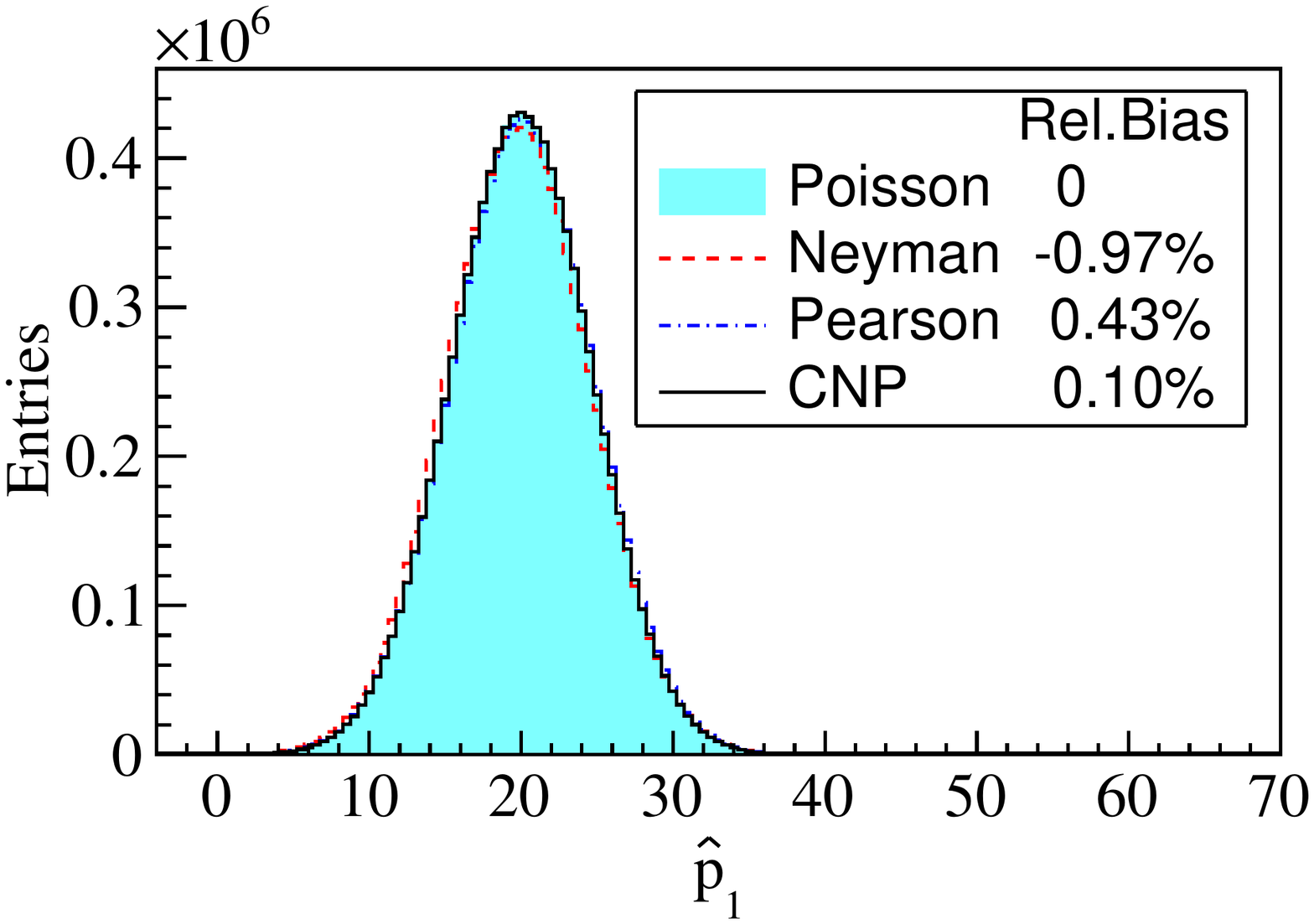}    
        \caption{Distributions of best-fit values of $p_0$ (left) and $p_1$ (right) for the example in \ref{sec:appendixE} using  $\chi^2_\mathrm{Poisson}$, $\chi^2_\mathrm{Neyman}$, $\chi^2_\mathrm{Pearson}$, and $\chi^2_\mathrm{CNP}$. The true values of $p_0$ and $p_1$ are 8 and 20, respectively. Ten million toy experiments are simulated. The numbers in the legend show the relative biases of the best-fit values.}
        \label{Fig::linear_comp}
    \end{center}
\end{figure}

Although in our examples in Sec.~\ref{sec:toy_model}, only one normalization parameter is considered (i.e.~the shape of the histogram is fixed), since the CNP chi-square is a better approximation to the Poisson-likelihood chi-square for counting statistics, we expect the improvement is general for any binned histograms with models including one or more parameters. Below we show an example where the shape of the histogram is linear, with the slope ($p_1$) and the y-intercept ($p_0$) being two free model parameters in the fit. The example is defined as follows:
\begin{equation}
n_i = p_0 + p_1  x_i \,,
\end{equation}
where $n_i$ is the number of counts in the i-th bin, and $x_i$ is the value of the bin center. $n_i$ is assumed to follow a Poisson distribution. 10 bins are considered in this example and $x_i$ ranges from 0.1 to 1 with a step of 0.1. The true values of $p_0$ and $p_1$ are assumed to be 8 and 20, respectively. 10 million toy experiments are generated according to this
setting. The distribution of best-fit values of $p_0$ and $p_1$ are shown in Fig.~\ref{Fig::linear_comp}. While the relative bias in $p_1$ (shape) is generally smaller than that of $p_0$ (normalization) given a chosen test statistic, the CNP chi-square yields smaller biases in both parameters as expected.


\bibliography{refs}
\bibliographystyle{JHEP}

\end{document}